\documentclass[12pt]{article} 
\pdfoutput=1
\usepackage{mathrsfs}
\usepackage{verbatim}
\usepackage{graphicx,epsf,rotate} 
\usepackage{cite,color,xspace}
\usepackage{amsmath}
\usepackage{amssymb}
\usepackage{comment}
\newcommand{\beq}{\begin{equation}}
\newcommand{\eeq}{\end{equation}}
\newcommand{\bea}{\begin{eqnarray}}
\newcommand{\eea}{\end{eqnarray}}
\newcommand{\ba}{\begin{array}}
\newcommand{\ea}{\end{array}}
\newcommand{\bec}{\begin{center}}
\newcommand{\eec}{\end{center}}
\newcommand{\bei}{\begin{itemize}}
\newcommand{\eei}{\end{itemize}}

\newcommand{\TeV}{\,\mathrm{TeV}}
\newcommand{\GeV}{\,\mathrm{GeV}}
\newcommand{\MeV}{\,\mathrm{MeV}}
\newcommand{\keV}{\,\mathrm{keV}}
\newcommand{\eV}{\,\mathrm{eV}}

\begin{document}
\thispagestyle{empty}

\begin{center}
{\Large \bf Displaced Vertex signatures of a pseudo-Goldstone} \\
\vspace{.3cm}
{\Large \bf sterile neutrino} \\
\vspace{1cm}
{\bf St\'ephane Lavignac$^{a,}$\footnote{\texttt{stephane.lavignac@ipht.fr}}
and Anibal D. Medina$^{b,}$\footnote{\texttt{anibal.medina@fisica.unlp.edu.ar}}} \\
\vspace{.5cm}
{\sl
\textit{$^{a}$Institut de Physique Th\'eorique, Universit\'e Paris Saclay, CNRS, CEA, \\
F-91191 Gif-sur-Yvette, France}\\
\textit{$^{b}$IFLP, CONICET - Dpto. de F\'{\i}sica, Universidad Nacional de La Plata, \\
C.C. 67, 1900 La Plata, Argentina}}
\end{center}

\vspace{.3cm}


\begin{abstract}

Low-scale models of neutrino mass generation often feature sterile neutrinos
with masses in the GeV-TeV range, which can be produced at colliders through
their mixing with the Standard Model neutrinos.
We consider an alternative scenario in which the sterile neutrino is produced
in the decay of a heavier particle, such that its production cross section
does not depend on the active-sterile neutrino mixing angles.
The mixing angles can be accessed through the decays of the sterile neutrino,
provided that they lead to observable displaced vertices.
We present an explicit realization of this scenario in which the sterile neutrino
is the supersymmetric partner of a pseudo-Nambu-Goldstone boson,
and is produced in the decays of higgsino-like neutralinos and charginos.
The model predicts the active-sterile neutrino mixing angles
in terms of a small number of parameters.
We show that a sterile neutrino with a mass between a few $10 \GeV$ and $200 \GeV$
can lead to observable displaced vertices at the LHC, and outline a strategy for
reconstructing experimentally its mixing angles.

\end{abstract} 


\def\thefootnote{\arabic{footnote}}
\setcounter{page}{0}
\setcounter{footnote}{0}

\newpage

\section{Introduction}     %

The existence of heavy sterile neutrinos is predicted by many models of neutrino mass generation,
such as the seesaw mechanism\cite{Minkowski:1977sc,Yanagida:1979as,GellMann:1980vs}
and its various incarnations.
In the standard, GUT-inspired picture, these heavy neutrinos have a mass $M \sim 10^{14} \GeV$
and order one Yukawa couplings $y$, thus providing a natural explanation for the smallness
of the Standard Model neutrino masses through the seesaw formula
$m_\nu \sim y^2 v^2 / M$.
However, there is no model-independent prediction for the masses of the sterile
neutrinos -- they could range from the GUT scale to the eV scale, or even below.
Of particular interest are sterile neutrinos with a mass between the GeV and the TeV scales,
which can lead to observable signatures at colliders 
(for reviews, see e.g. Refs.~\cite{Deppisch:2015qwa,Cai:2017mow}).
Most phenomenological studies and experimental searches assume a single
sterile neutrino $N$, which is produced through its mixing with the active (Standard Model) neutrinos,
parametrized by mixing angles $V_{N \alpha}$, where $\alpha = e, \mu, \tau$ is
the active lepton flavour.
Since the $V_{N \alpha}$ enter the sterile neutrino production cross section,
they can be measured directly, unless they are too small to give a detectable signal.

In this work, we consider an alternative production mechanism for the sterile neutrino
that does not depend on the active-sterile neutrino mixing angles.
Instead, the sterile neutrino is produced in the decay of a heavier particle,
whose production cross section is of typical electroweak size.
The mixing angles $V_{N \alpha}$ can be determined from the subsequent decays
of the sterile neutrino, provided that its total decay width is measured independently.
This can be done if the sterile neutrino decays are not prompt (as expected
if the active-sterile neutrino mixing is small),
in which case the decay width can be extracted from the distribution of displaced vertices.
We show that it is possible to probe values of the $V_{N \alpha}$
that would be out of reach if the production cross section were suppressed by the active-sterile
neutrino mixing, as usually assumed.
We study an explicit realization of this scenario in which the sterile neutrino is the
supersymmetric partner of the pseudo-Nambu Goldstone boson of a spontaneously
broken global $U(1)$ symmetry. This ``pseudo-Goldstone'' sterile neutrino is produced
in the decays of higgsino-like neutralinos and charginos and decays subsequently
via its mixing with the active neutrinos.
We outline an experimental strategy for measuring the active-sterile neutrino mixing
angles at the LHC, based on the reconstruction of final states involving displaced vertices.

The paper is organized as follows. In Section~\ref{sec:motivation}, after a brief summary
of current collider constraints, we introduce the sterile neutrino production mechanism
studied in this work and contrast it with the standard one.
In Section~\ref{sec:model}, we present an explicit model
in which the sterile neutrino is the supersymmetric partner of a
pseudo-Nambu Goldstone boson, and give its predictions for the active-sterile
neutrino mixing angles. In Section~\ref{sec:LHC_signatures}, we study the
experimental signatures of the model and discuss how the active-sterile
neutrino mixing angles could be reconstructed experimentally. We give our
conclusions in Section~\ref{sec:conclusion}.
Finally, Appendix~\ref{app:alignment} contains some technical details about the model
of Section~\ref{sec:model}, and Appendix~\ref{app:PNGB} discusses the
phenomenological constraints that apply to the pseudo-Nambu-Goldstone boson
whose supersymmetric partner is the sterile neutrino.

\section{Sterile neutrino production in heavy particle decays}   %
\label{sec:motivation}                                                                 %

Collider searches for heavy sterile neutrinos usually rely on a production mechanism
involving their mixing with the active neutrinos\footnote{Except for searches
performed in the framework of specific scenarios, like left-right symmetric extensions of the Standard Model,
in which the right-handed neutrinos are produced in decays of $W_R$ or $Z'$ gauge bosons~\cite{Keung:1983uu}.
In this case, the production cross section is controlled by the $SU(2)_R$ gauge coupling
and does not depend on the active-sterile neutrino mixing.}.
The active-sterile mixing angles can be measured directly, as they enter the production cross section.
Following this approach, early direct searches for heavy sterile neutrinos
were performed by the DELPHI experiment~\cite{Abreu:1996pa} at the LEP collider.
These searches were based on the production process $e^{+}e^{-}\to Z \to N \nu$,
followed by the decay $N \to l^\pm W^{\mp *}$ or $N \to \nu Z^*$.
Since the $Z N \nu_{\alpha}$ coupling is proportional to the active-sterile mixing angle
$V_{N\alpha}$ (where $\alpha=e,\mu$ or $\tau$), the production cross section
$\sigma (e^+ e^- \to N \nu)$ goes as $\sum_\alpha |V_{N\alpha}|^2$.
This allowed DELPHI to exclude mixing angles $|V_{N\alpha}|^2\, \gtrsim\, (2-3) \times 10^{-5}$
for sterile neutrino masses in the range $5 \GeV \lesssim m_N \lesssim 50 \GeV$,
independently of lepton flavour~\cite{Abreu:1996pa}. DELPHI was also able to set limits
on the $|V_{N\alpha}|^2$ for lower sterile neutrino masses,
using techniques involving displaced vertices.
For $|V_{Ne}|$, a stronger, indirect constraint can be derived
from the non-observation of neutrinoless double beta decay~\cite{Bamert:1994qh,Faessler:2014kka}.

At hadron colliders, the production process $p p (p \bar p) \to W^{\pm (*)} \to N l^\pm$,
followed by the decay $N \to  l^\pm W^{\mp (*)} \to l^\pm + 2\, \mbox{jets}$, leads to
the same-sign dilepton + 2 jets signature with no missing transverse energy characteristic
of a heavy Majorana neutrino~\cite{Keung:1983uu,Datta:1993nm}, which is essentially
free from Standard Model backgrounds. The $W$ boson from $N$ decay can also
go into a charged lepton and a neutrino, resulting in a trilepton signature.
Similarly to $Z N \nu_{\alpha}$, the $W^\pm N l^\mp_{\alpha}$ coupling is proportional
to $V_{N\alpha}$ and the production cross section $\sigma (p p \to N l^\pm_\alpha)$
goes as $|V_{N\alpha}|^2$. More precisely, the cross section for the process
$p p \to W^{\pm *} \to N l^\pm_\alpha$ mediated by $s$-channel $W$ boson exchange
is given by, at a center of mass energy $\sqrt{s}$~\cite{Pilaftsis:1991ug},
\begin{equation}
\sigma(s)\, = \int\! dx\! \int\! dy\, \sum_{q, \bar{q}'} \left[ f^{p}_{q}(x,Q^2)f^{p}_{\bar{q}'}(y,Q^2) \right]
  \hat{\sigma}(\hat{s})\, ,
\end{equation}
where $q=u, c$, $\bar{q}'=\bar{d}, \bar{s}$, $f^{p}_q$ is the parton distribution function
for the quark $q$ at $Q^2=\hat{s}=xys$, and $x$ and $y$ are the fractions of the proton
momentum carried by the interacting quark $q$ and antiquark $\bar q'$.
The parton subprocess cross section $\hat{\sigma}(\hat{s})$ is given by
\begin{equation}
\hat{\sigma}(\hat{s})\, =\, \frac{\pi \alpha^2_W}{72\, \hat{s}^2(\hat{s}-m^2_W)^2}\
  (\hat{s}-m^2_N)^2\, (2\hat{s}+m^2_N)\, |V_{N\alpha}|^2\, ,
\end{equation}
where $\alpha_W \equiv g^2/(4\pi)$, with $g$ the $SU(2)_L$ gauge coupling.
The collaborations ATLAS and CMS performed searches for heavy sterile neutrinos at $\sqrt{s} = 8 \TeV$,
using events with two jets and two leptons of the same charge, and set $m_N$-dependent
upper bounds on $|V_{N e}|^2$ and $|V_{N \mu}|^2$ in the mass range
$100 \GeV \leq m_N \leq 500 \GeV$ for ATLAS~\cite{Aad:2015xaa},
and $40 \GeV \leq m_N \leq 500 \GeV$ for CMS~\cite{Khachatryan:2015gha,Khachatryan:2016olu}.
The best limits were obtained by CMS, ranging from $|V_{N e}|^2\, \leq\, 1.5 \times 10^{-4}$,
$|V_{N \mu}|^2\, \leq\, 2 \times 10^{-5}$ for $m_N = 40 \GeV$
to $|V_{N e}|^2\, \leq\, 0.72$, $|V_{N \mu}|^2\, \leq\, 0.58$ for $m_N = 500 \GeV$, with
$|V_{N e}|^2\, \lesssim\, 10^{-2}$, $|V_{N \mu}|^2\, \lesssim\, 2 \times 10^{-3}$ around $100 \GeV$.
Using trilepton events with 35.9 fb$^{-1}$ of proton-proton collisions at $\sqrt{s}=13$ TeV,
CMS extended these constraints to the mass range $1 \GeV \leq m_N \leq 1.2 \TeV$,
providing upper bounds on $|V_{N e}|^2$ and $|V_{N \mu}|^2$ ranging from
$1.2 \times 10^{-5}$ to the unphysical value $1.8$, depending on $m_N$~\cite{Sirunyan:2018mtv}.
In particular, CMS slightly improved the DELPHI limits between $m_N = 10 \GeV$ and $50 \GeV$.
Using the same trilepton signature with 36.1 fb$^{-1}$ of proton-proton collisions
at $\sqrt{s}=13$ TeV, ATLAS obtained bounds similar to CMS in the mass range
$5 \GeV \leq m_N \leq 50 \GeV$, excluding $|V_{N e}|^2, |V_{N \mu}|^2\, \gtrsim 1.4 \times 10^{-5}$
between $20 \GeV$ and $30 \GeV$. By searching for displaced vertex signatures with a displacement
between $4$ and $300\, \mbox{mm}$, ATLAS was also able to probe values of $|V_{N \mu}|^2$
below $10^{-5}$ in the mass range $4.5 \GeV \leq m_N \leq 10 \GeV$, excluding
$|V_{N \mu}|^2\, \geq 1.5 \times 10^{-6}$ for $m_N \approx 9 \GeV$.

The possibility of probing smaller mixing angles via displaced vertex searches at the LHC
(allowing for larger displacements than in the ATLAS study just mentioned)
has been investigated in several phenomenological works~\cite{Helo:2013esa, Izaguirre:2015pga,
Gago:2015vma, Antusch:2017hhu, Cvetic:2018elt, Cottin:2018nms,  Abada:2018sfh,
Boiarska:2019jcw, Dib:2019ztn, Drewes:2019fou, Liu:2019ayx, Drewes:2019vjy, Jones-Perez:2019plk}
(see also Ref.~\cite{Alimena:2019zri} for a more general discussion about signatures
of long-lived particles at the LHC).
The requirement that the sterile neutrino does not decay promptly, while being
sufficiently produced through its mixing with the active neutrinos, restricts the
sensitivity of these searches to the low mass region, $m_N \lesssim (20 - 35) \GeV$,
depending on the available luminosity\footnote{Roughly speaking, the sterile neutrino
production cross section is proportional to $\sum_\alpha |V_{N\alpha}|^2$, while
its decay rate goes as $m^5_N \sum_\alpha |V_{N\alpha}|^2$. The requirement
that a significant number of sterile neutrinos are produced implies a lower bound
on $\sum_\alpha |V_{N\alpha}|^2$, while the requirement of displaced vertices
provides and upper bound on the combination $m^5_N \sum_\alpha |V_{N\alpha}|^2$.}.
For instance, Ref.~\cite{Drewes:2019fou} claims that displaced vertex searches
at the LHC could exclude values of $|V_{N e}|^2$ and $|V_{N \mu}|^2$ as small
as $5 \times 10^{-9}$ and masses up to $20 \GeV$ with an integrated luminosity
of $300\, \mbox{fb}^{-1}$, while the high luminosity LHC (HL-LHC) with an integrated luminosity
of $3\, \mbox{ab}^{-1}$ could probe $|V_{N e}|^2,\, |V_{N \mu}|^2 \approx 5 \times 10^{-10}$
and $m_N \approx 35 \GeV$. The sensitivity to $|V_{N \tau}|^2$ is typically smaller
by two orders of magnitude.

It is interesting to compare the above limits and sensitivities to the
predictions of typical GeV/TeV-scale seesaw models.
Using the na\"{i}ve seesaw formula $V_{N\alpha} \sim \sqrt{m_\nu/m_N}$,
one obtains $|V_{N\alpha}|^2 \sim 5 \times 10^{-12}$ for $m_N = 10 \GeV$,
and $|V_{N\alpha}|^2 \sim 10^{-13}$ for $m_N = 500 \GeV$, about seven orders
of magnitude below the best collider bounds.
Even displaced vertex searches at the LHC or HL-LHC do not seem to have
the potential to reach the vanilla seesaw model predictions and, moreover, are limited
to sterile neutrino masses below $40 \GeV$. All these searches are handicapped
by the fact that the sterile neutrino production cross section is suppressed by
the square of the active-sterile mixing. This prevents them from probing smaller
mixing angles and, in the case of displaced vertex searches, larger sterile neutrino masses.

In this paper, we consider the alternative possibility that the sterile neutrino production
mechanism does not depend on its mixing with active neutrinos.
In this case, the sensitivity of collider searches to small mixing angles is not limited
by the production rate, as the mixing only enters the sterile neutrino decays.
More specifically, we assume that the sterile neutrino is produced in the decays
of a heavier particle $\zeta$, whose production cross sections is of typical electroweak
size\footnote{Another possibility, namely the production of a pair of sterile neutrinos
in the decay of a $Z'$ gauge boson,
was considered in Refs.~\cite{Deppisch:2019kvs, Das:2019fee, Chiang:2019ajm}.}.
This requires that the sterile neutrino mixes with $\zeta$, in addition to its mixing
with active neutrinos. We further assume that $\zeta$ is produced in pairs, a property
encountered in many extensions of the Standard Model, where a parity symmetry
is often associated with the new particles. Sterile neutrinos are then produced as follows:
\begin{equation}
  p p \to \zeta \bar \zeta\, ,  \qquad  \zeta \to N + {\rm SM}\, ,
\end{equation}
where ``SM'' stands for Standard Model particles. If $m_{\zeta}\gg m_{N}$, these particles
are boosted and can be used as triggers for the signal of interest. Consider for instance
the production of two sterile neutrinos followed by their decay into a charged lepton
and two jets, $N \to l^\pm_\alpha W^{\mp (*)} \to l^\pm_\alpha q \bar q'$.
The rate for this process is given by
\begin{equation}
  \sigma (pp \to NN)\, {\rm BR}(N \to l_{\alpha}\, jj)\, {\rm BR}(N \to l_{\beta}\, jj)\,
    \propto\, \sigma (pp \to NN)\, \frac{|V_{N\alpha}|^2\ |V_{N \beta}|^2}{\Gamma_{N}^2}\, ,
\label{eq:NN_to_ll}
\end{equation}
where, assuming that the narrow width approximation is valid (i.e. $\Gamma_\zeta \ll m_\zeta$),
\begin{equation}
  \sigma (pp \to NN)\, =\, \sigma (pp \to \zeta \bar{\zeta}) \left[ {\rm BR}(\zeta \to N + {\rm SM}) \right]^2 .
\label{eq:NN_production}
\end{equation}
Eqs.~(\ref{eq:NN_to_ll}) and~(\ref{eq:NN_production}) clearly show that the active-sterile
mixing angles $V_{N \alpha}$ enter only the sterile neutrino decays, not its production.
This makes small mixing angles more easily accessible to collider searches than in the standard
scenario, in which the sterile neutrino is produced through its mixing with active neutrinos.
As can be seen from Eq.~(\ref{eq:NN_to_ll}), the number of events corresponding to a given
final state depends on the combinations $|V_{N \alpha}|^2 / \Gamma_N$.
An independent determination of $\Gamma_N$ is therefore needed in order to extract
the $V_{N \alpha}$ from experimental data. This can be done by measuring
the distribution of displaced vertices from $N$ decays, as we explain below.
Another virtue of displaced vertices is that they provide signals
which are essentially background free.
Indeed, Standard Model processes do not lead to displaced vertices (with the exception
of bottom and charm quarks, which produce small displacements and can be tagged).
This implies that a small number of events may be sufficient to measure the signal.

To conclude this section, let us explain how the sterile neutrino decay width
can be determined from the distribution of its displaced vertices.
The probability density for a particle travelling in a straight line to decay at a distance $r$
to a particular final state $i$ is given by
\begin{equation}
  P_i(r)\, =\, \frac{\Gamma_i}{\beta\gamma}\ e^{-\frac{\Gamma r}{\beta\gamma}}\, ,
\end{equation}
where $\Gamma_i$ is the corresponding decay rate, $\Gamma = \sum_j \Gamma_j$
the particle decay width, and we recall that $\beta \gamma = |\vec{p}|/m$,
with $\vec{p}$ the 3-momentum of the particle and $m$ its mass.
For an ensemble of identical particles, one needs to integrate over the particle momentum
distribution, on which $\beta\gamma$ depends.
However, to a good approximation, one can simply assume that all particles
have the same effective $(\beta \gamma)_{\rm eff}$, given by the peak value
of their $\beta\gamma$ distribution~\cite{Covi:2014fba}.
The number of particles decaying to the final state $i$ between $r_1$ and $r_2$, 
with $r_2>r_1$, is then given by 
\begin{equation}
  N_i(r_1,r_2)\, =\, N_0 \,\frac{\Gamma_i}{\Gamma} \left( e^{-\frac{\Gamma r_1}{(\beta \gamma)_{\rm eff}}}
    - e^{-\frac{\Gamma r_2}{(\beta \gamma)_{\rm eff}}} \right) ,
\label{eq:number_decay}
\end{equation}
where $N_0$ is the initial number of particles.
For two intervals $[r_1,r_2]$, $[r_3,r_4]$ such that $r_1 \ll r_2$ and $r_3 \ll r_4$, one can write
\begin{equation}
  \Gamma\, =\, \frac{(\beta \gamma)_{\rm eff}}{r_3 - r_1}\, \ln \left( \frac{N_i (r_1,r_2)}{N_i(r_3,r_4)} \right) .
\label{eq:Gamma}
\end{equation}
Thus, by measuring $(\beta \gamma)_{\rm eff}$ and the number of decays in two distance
intervals, one can obtain the decay width of the particle in the approximation described above.
More generally, i.e. without relying on the approximate formula~(\ref{eq:number_decay}),
the decay width can be extracted from the shape of the distribution of displaced vertices
corresponding to a given final state, provided that the mass and the momentum distribution
of the particle (hence its $\beta \gamma$ distribution) can be reconstructed experimentally.

\section{An explicit model: sterile neutrino as the supersymmetric partner of a pseudo-Nambu-Goldstone boson}    %
\label{sec:model}                                                                                                                                                           %

We now present an explicit realization of the scenario discussed in the previous section.
The sterile neutrino is identified with the supersymmetric partner of a pseudo-Nambu-Goldstone
boson\footnote{For earlier realizations of this idea, in which the sterile neutrino was assumed
to be light, see Refs.~\cite{Chun:1995js,Chun:1995bb,Chun:1999kd,Choi:2001cm}.} (PNGB)
and mixes both with active neutrinos and higgsinos. Its mixing with higgsinos is relatively large,
such that it is predominantly produced in neutralino and chargino decays.

\subsection{The model}
\label{subec:model}

The model we consider is an extension of the Minimal Supersymmetric Standard Model (MSSM)
with a global $U(1)$ symmetry under which the lepton and Higgs fields (but not the quark fields)
are charged. We assume that this symmetry is spontaneously broken at some high scale $f$
by the vacuum expectation value (VEV) of a scalar field belonging to a chiral superfield $\Phi$
with charge $-1$ (we also assume a small source of explicit breaking to avoid a massless Goldstone boson).
The charges of the superfields $L_i$, $\bar e_i$, $H_u$
and $H_d$ are denoted by $l_i$, $e_i$, $h_u$ and $h_d$, respectively. We choose the
symmetry to be generation-independent and vector-like, i.e. $e_i = - l_i \equiv - l$, in order to avoid
dangerous flavor-changing processes~\cite{Wilczek:1982rv} and astrophysical constraints on
the pseudo-Nambu-Goldstone boson $a$~\cite{Viaux:2013lha} (see Appendix~\ref{app:PNGB} for details).
We further assume $h_u = 0$, so that the top quark Yukawa coupling is invariant under
the global symmetry and therefore unsuppressed by powers of the symmetry breaking parameter.

We are then left with two independent charges $l$ and $h_d$, which we assume to be positive integers.
With this choice, the down-type quark and charged lepton Yukawa couplings, as well
as the $\mu$-term, are not allowed by the global symmetry and must arise
from higher-dimensional superpotential operators involving the field $\Phi$:
\begin{eqnarray}
  W\! & =\! & \kappa_0\, H_u H_d \Phi \left( \frac{\Phi}{M} \right)^{h_d-1}\! +\, \kappa_i\, H_u L_i \Phi \left( \frac{\Phi}{M} \right)^{l-1}\!
    -\, y^e_{ij}\, L_i \bar e_j H_d \left( \frac{\Phi}{M} \right)^{h_d}
    -\, y^d_{ij}\, Q_i \bar d_j H_d \left( \frac{\Phi}{M} \right)^{h_d}  \nonumber \\
  && +\ \lambda^u_{ij}\, Q_i \bar u_j H_u + \frac{1}{2}\, y_{ijk}\, L_i L_j \bar e_k \left( \frac{\Phi}{M} \right)^{l}
    +\, y'_{ijk}\, L_i Q_j \bar d_k \left( \frac{\Phi}{M} \right)^{l}\, ,
\label{eq:W_Phi}
\end{eqnarray}
where $M \gg f \equiv \langle \Phi \rangle$ is the scale of the new physics that generates these operators.
In the superpotential~(\ref{eq:W_Phi}), we included terms that lead to $R$-parity violating interactions,
with the exception of the baryon number violating couplings $\bar u_i \bar d_j \bar d_k$,
which we assume to be forbidden by some symmetry, such as the $\mathbb{Z}_3$ baryon
parity of Ref.~\cite{Ibanez:1991pr}.
In principle, Eq.~(\ref{eq:W_Phi}) should also contain a term
$\frac{1}{4}\, \kappa_{ij}\, L_i L_j H_u H_u \Phi^{2l} / M^{2l+1}$, which after spontaneous symmetry breaking
induces the Weinberg operator $L_i L_j H_u H_u$. We will discuss this contribution
to neutrino masses later in this section.

The spontaneous breaking of the global symmetry generates the following
superpotential\footnote{In going from Eq.~(\ref{eq:W_Phi}) to Eq.~(\ref{eq:W_Phi_hat}), we dropped
the terms involving more than one power of $\hat \Phi$, since, in addition to being suppressed
by powers of the large scale $f$,
they do not contribute to the mixing between the pseudo-Goldstone fermion and the MSSM fermions.
As we are going to see, it is this mixing that determines the collider phenomenology of the sterile neutrino.}:
\begin{eqnarray}
  W\! & =\! & \hat \mu_0\, H_u H_d + \hat \mu_i\, H_u L_i + \hat \lambda_0\, H_u H_d \hat \Phi + \hat \lambda_i\, H_u L_i \hat \Phi
    - \hat \lambda^e_{ij}\, L_i \bar e_j H_d - \hat \lambda^d_{ij}\, Q_i \bar d_j H_d  \nonumber \\
  && +\, \lambda^u_{ij}\, Q_i \bar u_j H_u + \frac{1}{2}\, \hat \lambda_{ijk}\, L_i L_j \bar e_k
    + \hat \lambda'_{ijk}\, L_i Q_j \bar d_k\, ,
\label{eq:W_Phi_hat}
\end{eqnarray}
where $\hat \Phi$ stands for the shifted superfiel $\Phi - f$ and
\begin{equation}
  \begin{array}{llll} \hat \mu_0\, =\, \kappa_0 f \epsilon^{h_d - 1}\, , \quad & \hat \lambda_0\, =\, h_d \kappa_0 \epsilon^{h_d-1}\,
    =\, h_d \hat \mu_0 / f\, , \quad & \hat \lambda^e_{ij}\, =\, y^e_{ij} \epsilon^{h_d}\, , \quad
    & \hat \lambda^d_{ij}\, =\, y^d_{ij} \epsilon^{h_d}\, ,  \\
  \hat \mu_i\, =\, \kappa_i f \epsilon^{l - 1}\, , \quad & \hat \lambda_i\, =\, l \kappa_i \epsilon^{l-1}\, =\, l \hat \mu_i / f\, , \quad 
    & \hat \lambda_{ijk}\, =\, y_{ijk} \epsilon^{l}\, , \quad & \hat \lambda'_{ijk}\, =\, y'_{ijk} \epsilon^{l}\, ,  \end{array} 
\label{eq:initial_parameters}
\end{equation}
in which $\epsilon \equiv \langle \Phi \rangle / M = f / M \ll 1$.
We will assume $l > h_d$, such that the $R$-parity violating parameters $\hat \mu_i$, $\hat \lambda_i$,
$\hat \lambda_{ijk}$ and $\hat \lambda'_{ijk}$ are suppressed by a factor $\epsilon^{l-h_d}$
relative to the corresponding $R$-parity conserving parameters:
\begin{equation}
  \hat \mu_i\, \sim\, \hat \mu_0\, \epsilon^{l-h_d}\, ,
    \qquad \hat \lambda_i\, \sim\, \hat \lambda_0\, \epsilon^{l-h_d}\, ,
    \qquad \hat \lambda_{ijk}\, \sim\, \hat \lambda^e_{jk}\, \epsilon^{l-h_d}\, ,
    \qquad \hat \lambda'_{ijk}\, \sim\, \hat \lambda^d_{jk}\, \epsilon^{l-h_d}\, ,
\label{eq:hierarchy_initial_parameters}
\end{equation}
where, for definiteness, we have assumed $\kappa_i \sim \kappa_0$, $y_{ijk} \sim y^e_{jk}$
and $y'_{ijk} \sim y^d_{jk}$.
$R$-parity violation is therefore automatically suppressed by the choice of the $U(1)$ charges; there is
no need to invoke an ad hoc hierarchy between $R$-parity even and $R$-parity odd coefficients
in the superpotential~(\ref{eq:W_Phi}).
Note that the coefficicents $y^e_{ij}$, $y^d_{ij}$ and $\lambda^u_{ij}$ must have a hierarchical
flavour structure in order to account for the fermion mass spectrum
(this cannot be explained by the $U(1)$ symmetry itself, since it is generation independent).

The chiral superfield $\hat \Phi$ contains the pseudo-Nambu-Goldstone boson and its supersymmetric partners.
It can be written as:
\begin{equation}
  \hat \Phi\, =\, \frac{s+i a}{\sqrt{2}}\, + \sqrt{2}\, \theta \chi + \theta^2 F\, ,
\end{equation}
where $a$ is the PNGB, $s$ its scalar partner, which is assumed to get
a large mass from supersymmetry breaking, and $\chi$ its fermionic partner (hereafter referred
to as the pseudo-Goldstone fermion or sterile neutrino), whose mass $m_\chi$ also predominantly
arises from supersymmetry breaking. In particular, $m_\chi$ receives an irreducible contribution
proportional to the gravitino mass~\cite{Cheung:2011mg}.
By contrast, the pseudo-Nambu-Goldstone boson $a$ only obtains its mass from the sources of explicit
global symmetry breaking, assumed to be small.
The hierarchy of mass scales is therefore $m_a \ll m_\chi \ll m_s$, and we will consider values
of $m_\chi$ in the few 10 GeV to few 100 GeV range in the following.
As discussed in Appendix~\ref{app:PNGB}, $m_a$ is constrained to be larger than about $400 \MeV$
by cosmological and astrophysical observations.

Before we can derive the interactions of the pseudo-Goldstone fermion,
we must take into account the effect of supersymmetry breaking.
Since $R$-parity has not been imposed, the scalar potential includes soft supersymmetry
breaking terms that violate $R$-parity, which in turn induce vevs for the sneutrinos.
After redefining the superfields $H_d$ and $L_i$ in such a way that
{\it (i)}~only the scalar component of $H_d$ gets a vev
and {\it (ii)} charged lepton Yukawa couplings are diagonal,
we end up with the following superpotential (see Appendix~\ref{app:alignment} for details):
\begin{eqnarray}
  W\! & =\! & \mu_0\, H_u H_d + \mu_i\, H_u L_i + \lambda_0\, H_u H_d \hat \Phi + \lambda_i\, H_u L_i \hat \Phi
    +\, \cdots\ ,
\label{eq:superpotential}
\end{eqnarray}
where $H_d$ and $L_i$ are now the physical down-type Higgs and lepton doublet superfields.
We have dropped the Yukawa couplings and the trilinear $R$-parity violating couplings $\lambda_{ijk}$
and $\lambda'_{ijk}$, as they do not contribute to the mixing of the pseudo-Goldstone fermion
with other fermions. As shown in Appendix~\ref{app:alignment}, the parameters $\mu_0$,
$\mu_i$, $\lambda_0$ and $\lambda_i$ can be written as
\begin{equation}
  |\mu_0| = \mu \sqrt{1 - \xi^2}\, \simeq \mu\, ,  \qquad  \mu_i = c_i \mu\, \xi\, , \qquad
    \lambda_0 \simeq h_d\, \frac{\mu}{f}\, ,  \qquad  \lambda_i = d_i \frac{\mu}{f}\, \xi\, ,
\label{eq:parameters}
\end{equation}
where $\mu \equiv \sqrt{|\hat \mu_0|^2 + \sum_i |\hat \mu_i|^2}$, $\xi \sim \epsilon^{l-h_d}$
is an overall measure of bilinear $R$-parity violation defined in Appendix~\ref{app:alignment},
and  $c_i$, $d_i$ are order one coefficients (with $\sum_i |c_i|^2 = 1$).
We thus have the order-of-magnitude relations
\begin{equation}
  \frac{\mu_i}{\mu_0}\ \sim\ \frac{\lambda_i}{\lambda_0}\ \sim\ \xi\ \sim\ \epsilon^{l-h_d}\, , \qquad
  \lambda_0\ \sim\ \frac{\mu}{f}\ .
\end{equation}
Since $\xi$ and $\mu/f$ are small quantities, this implies $\mu_i \ll \mu_0$ and $\lambda_i \ll \lambda_0 \ll 1$.

\subsection{Sterile neutrino interactions}
\label{subec:mixing}

The superpotential~(\ref{eq:superpotential}) includes terms mixing the pseudo-Goldstone
fermion $\chi$ with leptons, promoting it to a sterile neutrino. Since it is a gauge singlet,
its interactions arise from its mixing with the other neutral fermions, namely the active neutrinos
$\nu_i$, the neutral higgsinos $\tilde h^0_u$, $\tilde h^0_d$ and the gauginos $\lambda_\gamma$,
$\lambda_Z$. This mixing is encoded in the ($8 \times 8$) neutralino mass matrix, which
in the 2-component fermion basis
$\psi^0 = (\lambda_\gamma, \lambda_Z, \tilde h^0_u, \tilde h^0_d, \nu_i, \chi)$ is given by
\begin{equation}
  M_N\, =\, \left( \begin{array}{cccccc}
    c^2_W M_1 + s^2_W M_2 & c_W s_W (M_2 - M_1) & 0 & 0 & 0_{1 \times 3} & 0  \\
    c_W s_W (M_2 - M_1) & s^2_W M_1 + c^2_W M_2 & - \frac{g v_u}{\sqrt{2} c_W} & \frac{g v_d}{\sqrt{2} c_W} & 0_{1 \times 3} & 0  \\
    0 & - \frac{g v_u}{\sqrt{2} c_W} & 0 & - \mu_0 & - \mu_j & - \lambda_0 v_d  \\
    0 & \frac{g v_d}{\sqrt{2} c_W} & - \mu_0 & 0 & 0_{1 \times 3} & - \lambda_0 v_u  \\
    0_{3 \times 1} & 0_{3 \times 1} & - \mu_i & 0 & \delta (M_\nu)_{ij} & - \lambda_i v_u  \\
    0 & 0 & - \lambda_0 v_d & - \lambda_0 v_u & - \lambda_j v_u & m_\chi
  \end{array} \right) ,
\label{eq:M_N}
\end{equation}
where $c_W \equiv \cos \theta_W$, $s_W \equiv \sin \theta_W$, $\tan \beta \equiv v_u / v_d$
and $v_u$, $v_d$ are the VEVs of the two Higgs doublets of the supersymmetric Standard Model.
The $3 \times 3$ matrix
$\delta M_\nu$ contains small contributions to the $\nu_i$ mass terms arising from loops
induced by the trilinear $R$-parity violating couplings $\lambda_{ijk}$ and $\lambda'_{ijk}$,
as well as from high-scale physics parametrized by the superpotential operators
$L_i L_j H_u H_u \Phi^{2l} / M^{2l+1}$.
The chargino mass matrix is given by, in the bases
$\psi^- = (\lambda^-, \tilde h^-_d, l^-_i)$ and $\psi^+ = (\lambda^+, \tilde h^+_u, \bar e_j)$,
\begin{equation}
  M_C\, =\, \left( \begin{array}{ccc}
    M_2 & g v_u & 0_{1 \times 3}  \\
    g v_d & \mu_0 & 0_{1 \times 3}  \\
    0_{3 \times 1} & \mu_i & m_{l_i} \delta_{ij}
  \end{array} \right) ,
\label{eq:M_C}
\end{equation}
where $m_{l_i}$ are the charged lepton masses.
The neutralino and chargino mass matrices are diagonalized by unitary matrices $N$, $V$ and $U$,
which relate the mass eigenstates $\chi^0_i$ and $\chi^+_i$ to the gauge eigenstates
$\psi^0_i$ and $\psi^+_i$:
\begin{equation}
  \chi^0_i = N_{ij} \psi^0_j\, \quad (i, j = 1 \dots 8),  \qquad
    \chi^+_i = V_{ij} \psi^+_j , \quad  \chi^-_i = U_{ij} \psi^-_j  \quad  (i, j = 1 \dots 5)\, .
\label{eq:mass_eigenstates}
\end{equation}

The neutralino mass matrix~(\ref{eq:M_N}) has a ``seesaw'' structure, with the upper left $4 \times 4$
block (associated with the MSSM neutralinos) containing the largest entries, while the elements
of the off-diagonal block are suppressed by $\mu_i / \mu_0 \sim \xi$ or $\lambda_0 \sim \mu/f$,
and the lower right $4 \times 4$ block has $m_\chi\, (\ll \mu_0, M_1, M_2)$ as its largest entry,
$\delta (M_\nu)_{ij}$ and $\lambda_i v_u \sim \xi (\mu/f) v_u$ being much smaller.
Similarly, the chargino mass matrix~(\ref{eq:M_C}) has a dominant $2 \times 2$ upper left block
corresponding to the MSSM charginos. This results in the following mass spectrum:
\begin{equation}
  m_{\tilde \chi^0_{1,2,3}} \ll\, m_{\tilde \chi^0_4}\, \ll\, m_{\tilde \chi^0_{5,6,7,8}}\, , \qquad
  m_{\tilde \chi^\pm_{1,2,3}} \ll\, m_{\tilde \chi^\pm_{4,5}}\, ,
\label{eq:ino_spectrum}
\end{equation}
where $\tilde \chi^0_{1,2,3}$ and $\tilde \chi^\pm_{1,2,3}$ can be identified with the
active neutrinos and charged leptons, respectively, $\tilde \chi^0_4$ is the sterile neutrino,
and $\tilde \chi^0_{5,6,7,8}$, $\tilde \chi^\pm_{4,5}$ are mostly the MSSM neutralinos
and charginos. We therefore rename the mass eigenstates~(\ref{eq:ino_spectrum})
in the following way:
\begin{equation}
  \nu_{1,2,3} \equiv \tilde \chi^0_{1,2,3}\ ,  \quad  N \equiv \tilde \chi^0_4\ , \quad 
    \tilde \chi^0_{1,2,3,4} \equiv \tilde \chi^0_{5,6,7,8}\ ,  \qquad
  l^\pm_{1,2,3} \equiv \tilde \chi^\pm_{1,2,3}\ ,  \quad  \tilde \chi^\pm_{1,2} \equiv \tilde \chi^\pm_{4,5}\ .
\label{eq:ino_spectrum2}
\end{equation}
Due to the specific hierarchical structure of the neutralino mass matrix, the sterile neutrino
mass is approximately given by $m_\chi$ ($m_N \simeq m_\chi$).
The hierarchy among the entries of $M_N$ and $M_C$ also implies that the mixing
between states well separated in mass is small.
As can be seen from Eq.~(\ref{eq:M_C}), the mixing between charginos and charged leptons
is suppressed by $\mu_i / \mu_0 \sim \xi$, while the mixing between the sterile neutrino, neutrinos
and neutralinos has a more complicated structure and depends on the small parameters
$\mu_i / \mu_0 $, $\lambda_0$ and $\lambda_i$.
These mixings induce new interactions between gauge bosons and fermions that are
absent in the Standard Model. We will be interested in the ones that are relevant
for the production and decay of the sterile neutrino, namely (assuming $\mu \simeq |\mu_0| \ll M_1, M_2$,
such that the lightest neutralinos $\tilde \chi^0_{1,2}$ and chargino $\tilde \chi^\mp_1$ are mainly higgsinos):
\begin{equation}
  Z \tilde \chi^0_{1,2} N\, ,  \quad  Z N \nu_i\, ,  \quad
  W^\pm \tilde \chi^\mp_1 N\, ,  \quad  W^\pm N l^\mp_i\, .
\end{equation}
All these couplings are suppressed by small mixing angles.

\subsection{Constraints from neutrino data and active-sterile mixing angles}
\label{subsec:VNa}

The model contains a large number of parameters (in addition to the supersymmetric
parameters $\mu$, $M_1$, $M_2$ and $\tan \beta$, the chargino and neutralino
mass matrices depend on $\lambda_0$, $\lambda_i$ and $\mu_i$ ($i = 1,2,3$), or alternatively
on $\lambda_0$, $c_{1,2}$, $d_{1,2,3}$ and $\xi$).
However, the requirement that it should be consistent with neutrino oscillation data
fixes a lot of them. An efficient way of taking this constraint into account is to express
the model parameters in terms of the neutrino parameters $\Delta m^2_{31}$,
$\Delta m^2_{21}$ and $U_{\alpha i}$ (where $U$ denotes the PMNS matrix, which
controls flavour mixing in the lepton sector).
In order to do this, we take advantage of the strongly hierarchical structure of the neutralino
mass matrix~(\ref{eq:M_N}) to derive the active neutrino mass matrix in the
seesaw approximation:
\begin{equation}
  (M_\nu)_{\alpha\beta}\, \simeq\,  (\delta M_\nu)_{\alpha\beta}
    - A (c_\alpha - d_\alpha / h_d) (c_\beta - d_\beta / h_d) - B c_\alpha c_\beta\, ,
\label{eq:effective_Mnu}
\end{equation}
where $A \equiv \lambda^2_0 v^2_u\, \xi^2 / m_\chi$,
$B \equiv (c^2_W M_1 + s^2_W M_2) m^2_Z \cos^2 \beta\, \xi^2 / (M_1 M_2)$,
and the lepton family indices have been renamed from $i = 1,2,3$ to $\alpha = e, \mu, \tau$
to stress that we are working in the charged lepton mass eigenstate basis.
For simplicity, we assume $\delta M_\nu = 0$ in the following, since its entries
are small\footnote{The superpotential terms
$\frac{1}{4}\, \kappa_{\alpha \beta}\, L_\alpha L_\beta H_u H_u \Phi^{2l} / M^{2l+1}$ give a contribution
$(\delta M_\nu)_{\alpha \beta} = \frac{1}{2}\, \kappa_{\alpha \beta}\, \epsilon^{2l+1} v^2_u / f$ 
to the neutrino mass matrix. For the values of the model parameters considered in this paper:
$\epsilon = 0.1$, $l = 6$ and $f = 15.8 \TeV / \sqrt{m_N / 100 \GeV}$, this gives
$(\delta M_\nu)_{\alpha \beta} \sim (10^{-4} \eV)\, \kappa_{\alpha \beta}$, which is
too small to affect significantly the neutrino oscillation parameters.
As for the lepton-slepton and quark-squark loops induced by the trilinear
$R$-parity violating couplings $\lambda$ and $\lambda'$, they are suppressed by $(\lambda)^2$
and $(\lambda')^2$, respectively, which are smaller than $\mathcal{O} (\epsilon^{2l})$,
as shown in Appendix~\ref{app:alignment}.}
compared with the values $(M_\nu)_{\alpha \beta} \gtrsim (0.001 \div 0.05) \eV$
suggested by neutrino oscillation data.
With this choice, only two neutrinos become massive\footnote{This can already be seen
at the level of Eq.~(\ref{eq:M_N}), whose determinant vanishes for $\delta M_\nu = 0$.}.
We further assume that the neutrino mass ordering is normal, i.e. $m_1 = 0$.
Then the neutrino oscillation parameters are reproduced by the mass matrix~(\ref{eq:effective_Mnu}) with
\begin{equation}
  A = m_3\, , \quad B = m_2\, , \quad c_\alpha = U^*_{\alpha 2}\, , \quad d_\alpha = h_d (U^*_{\alpha 2} - U^*_{\alpha 3})\, ,
\label{eq:maximal_mixing}
\end{equation}
where $m_3 = \sqrt{\Delta m^2_{31}}$ and $m_2 = \sqrt{\Delta m^2_{21}}$.
This fixes the parameters $c_\alpha$ and $d_\alpha$, as well as $\xi$ (as a function
of $M_1$, $M_2$ and $\tan \beta$) and
$\lambda_0$ (as a function of $m_\chi \simeq m_N$, $M_1$, $M_2$ and $\tan \beta$).
With the additional input of $\mu$ and the help of Eq.~(\ref{eq:parameters}), the parameters
$\mu_\alpha$ and $\lambda_\alpha$ can be reconstructed.
Thus, for a given set of neutrino parameters, the model has only five free parameters
in the limit $\delta M_\nu = 0$: $\mu$, $M_1$, $M_2$, $\tan \beta$ and $m_N$.
For the reference values $\mu = 500 \GeV$, $M_1 = 1 \TeV$, $M_2 = 2 \TeV$
and $\tan \beta = 10$ used in Section~\ref{sec:LHC_signatures}, one obtains
$\xi = 1.29 \times 10^{-5}$ and $\lambda_0 = 3.16 \times 10^{-2} \sqrt{m_N / 100 \GeV}$.
Given the order-of-magnitude relation $\xi \sim \epsilon^{l-h_d}$, this value of $\xi$
is consistent with the choice of ``fundamental'' parameters
$\epsilon = 0.1$, $l = 6$ and $h_d = 1$, which we adopt from now on.
The value of $\lambda_0$ corresponds to a global symmetry breaking scale
$f = h_d\, \mu / \lambda_0 = 15.8 \TeV / \sqrt{m_N / 100 \GeV}$.

Having traded some of the model parameters for the neutrino oscillation parameters,
we can now derive a simple expression for the active-sterile neutrino mixing angles
(still in the seesaw approximation):
\begin{equation}
  V_{N \alpha}\, \simeq\, \sqrt{\frac{m_3}{m_N}}\ U^*_{\alpha 3}\, .
\label{eq:V_N_alpha_max}
\end{equation}
For the current best fit values of the oscillation parameters~\cite{Esteban:2020cvm,deSalas:2020pgw},
this gives
\begin{equation}
  (|V_{N e}|, |V_{N \mu}|, |V_{N \tau}|)\ \simeq\ (1.1, 5.3, 4.6) \times 10^{-7} \left( {\frac{100\, \GeV}{m_N}} \right)^{1/2}\, .
\end{equation}
In the approximation in which we are working, where all states heavier than the sterile neutrino
are decoupled, these mixing angles enter the vertices $W^\pm N l^\mp_\alpha$ and $Z N \nu_\alpha$,
i.e. the corresponding Lagrangian terms are
$(g V_{N \alpha} / 2 c_W) Z_\mu \bar N \gamma^\mu \nu_\alpha
+ (g V_{N \alpha} / \sqrt{2}) \left( W^+_\mu \bar N \gamma^\mu \ell^-_\alpha + \mbox{h.c.} \right)$.
We checked numerically that this gives a very good approximation to the exact $W^\pm N l^\mp_\alpha$
couplings, obtained by expressing the $W$ boson--fermion interactions in terms of the neutralino
and chargino mass eigenstates.
The approximation is less reliable for the individual  $Z N \nu_\alpha$ couplings, but becomes
very good after summing over the neutrino flavours.

In fact, Eq.~(\ref{eq:maximal_mixing}) is not the most general solution to Eq.~~(\ref{eq:effective_Mnu}).
Another solution is
\begin{equation}
  A = m_2\, , \quad B = m_3\, , \quad c_\alpha = U^*_{\alpha 3}\, , \quad d_\alpha = h_d (U^*_{\alpha 3} - U^*_{\alpha 2})\, ,
\label{eq:minimal_mixing}
\end{equation}
giving
\begin{equation}
  V_{N \alpha}\, \simeq\, \sqrt{\frac{m_2}{m_N}}\ U^*_{\alpha 2}\, .
\label{eq:V_N_alpha_min}
\end{equation}
It is not difficult to show that the general solution is of the form
\begin{equation}
  V_{N \alpha}\ \simeq\ R_{11}\, \sqrt{\frac{m_3}{m_N}}\, U^*_{\alpha 3}\,
    +\, R_{12}\, \sqrt{\frac{m_2}{m_N}}\, U^*_{\alpha 2}\, ,
\label{eq:V_N_alpha_general}
\end{equation}
where $R$ is a $2 \times 2$ complex orthogonal matrix\footnote{Thus, in full generality,
the model has seven free parameters in the limit $\delta M_\nu = 0$: $\mu$, $M_1$,
$M_2$, $\tan \beta$, $m_N$ and a complex parameter parametrizing the matrix $R$
(strictly speaking, the parametrization of $R$ also involves a sign distinguishing
between $\det R = + 1$ and $\det R = - 1$).}.
We will refer to Eq.~(\ref{eq:maximal_mixing})
and Eq.~(\ref{eq:minimal_mixing}) as the ``maximal mixing'' and ``minimal mixing'' solutions,
respectively, even though larger values of $\sum_\alpha |V_{N \alpha}|^2$ can be obtained
for complex $R_{11}$ and $R_{12}$.

Approximate analytic expressions for the other relevant mixing angles can be obtained in the same way,
diagonalizing the $8 \times 8$ neutralino mass matrix by blocks like in the seesaw approximation,
and further assuming $\mu \ll M_1, M_2$ (so that
$\chi^0_1 \simeq (\tilde h^0_u - \tilde h^0_d) / \sqrt{2}$ and
$\chi^0_2 \simeq i (\tilde h^0_u + \tilde h^0_d) / \sqrt{2}$\, ).
With a very good numerical accuracy, the mixing between the active neutrinos and the
mostly-higgsino neutralinos $\tilde \chi^0_{1,2}$ is given by, in the maximal mixing case
\begin{equation}
  V_{\tilde \chi^0_1 \alpha}\, \simeq\, - \frac{1}{\sqrt{2}}\, U^*_{\alpha 2}\, \xi
    \left( 1 + \frac{m_3}{\mu\, \xi^2}\, \frac{U^*_{\alpha 3}}{U^*_{\alpha 2}}\, (1 + \cot \beta) \right) ,
\label{eq:V_chi01_alpha_max}
\end{equation}
\begin{equation}
  V_{\tilde \chi^0_2 \alpha}\, \simeq\, \frac{i}{\sqrt{2}}\, U^*_{\alpha 2}\, \xi
    \left( 1 - \frac{m_3}{\mu\, \xi^2}\, \frac{U^*_{\alpha 3}}{U^*_{\alpha 2}}\, (1 - \cot \beta) \right) .
\label{eq:V_chi02_alpha_max}
\end{equation}
In practice, the second term in the parenthesis can be neglected as long as $\xi \gtrsim 10^{-6}$.
In the general case, Eqs.~(\ref{eq:V_chi01_alpha_max})--(\ref{eq:V_chi02_alpha_max})
are replaced by (dropping the second term)
\begin{equation}
  V_{\tilde \chi^0_1 \alpha}\, \simeq\, - \left( R_{21}\, \sqrt{\frac{m_3}{m_{\rm eff}}}\, U^*_{\alpha 3}
    + R_{22}\, \sqrt{\frac{m_2}{m_{\rm eff}}}\, U^*_{\alpha 2} \right) ,
  \qquad \quad  V_{\tilde \chi^0_2 \alpha}\, \simeq\, - i\, V_{\tilde \chi^0_1 \alpha}\, ,
\label{eq:V_chi01_alpha_general}
\end{equation}
where $m_{\rm eff} \equiv 2 (c^2_W M_1 + s^2_W M_2) m^2_Z \cos^2 \beta / (M_1 M_2)$.
Finally, the mixing between the sterile neutrino and the mostly-higgsino neutralinos
$\tilde \chi^0_{1,2}$ is given by
\begin{equation}
  V_{\tilde \chi^0_1 N}\, \simeq\, \frac{\lambda_0}{\sqrt{2}}\, \frac{v \sin \beta}{\mu}
    \left( 1 + \cot \beta - \frac{(c^2_W M_1 + s^2_W M_2) m^2_Z}{\mu M_1 M_2} \right)
    \approx\, \frac{\lambda_0}{\sqrt{2}}\, \frac{v \sin \beta}{\mu}\ ,
\label{eq:V_chi01_N}
\end{equation}
\begin{equation}
  V_{\tilde \chi^0_2 N}\, \simeq\, i\, \frac{\lambda_0}{\sqrt{2}}\, \frac{v \sin \beta}{\mu}
    \left( 1 - \cot \beta + \frac{(c^2_W M_1 + s^2_W M_2) m^2_Z}{\mu M_1 M_2} \right)
    \approx\, i\, \frac{\lambda_0}{\sqrt{2}}\, \frac{v \sin \beta}{\mu}\ .
\label{eq:V_chi02_N}
\end{equation}
We checked numerically that Eqs.~(\ref{eq:V_chi01_alpha_max})--(\ref{eq:V_chi02_alpha_max})
and~(\ref{eq:V_chi01_N})--(\ref{eq:V_chi02_N}) provide
good approximations
for the mixing angles appearing at the $Z \tilde \chi^0_{1,2} \nu_\alpha$ and
$Z \tilde \chi^0_{1,2} N / W^\pm \tilde \chi^\mp_1 N$ vertices, respectively.

\section{Collider signatures of the pseudo-Goldstone sterile neutrino}   %
\label{sec:LHC_signatures}                                                                     %

We are now ready to study the collider signatures of the pseudo-Goldstone sterile neutrino
described in Section~\ref{sec:model}, focusing on the LHC. We will show in particular
that if the mass of the sterile neutrino is around 100 GeV, most of its decays occur
within the ATLAS and CMS detectors and lead to displaced vertices.
Assuming that the events can be reconstructed efficiently, we discuss how the mass
of the sterile neutrino and its mixing angles can be determined from the experimental data.

\subsection{Model parameters and mixing angles}
\label{subsec:parameters}

\begin{figure}[t]
\centering
\includegraphics[width=0.7\textwidth]{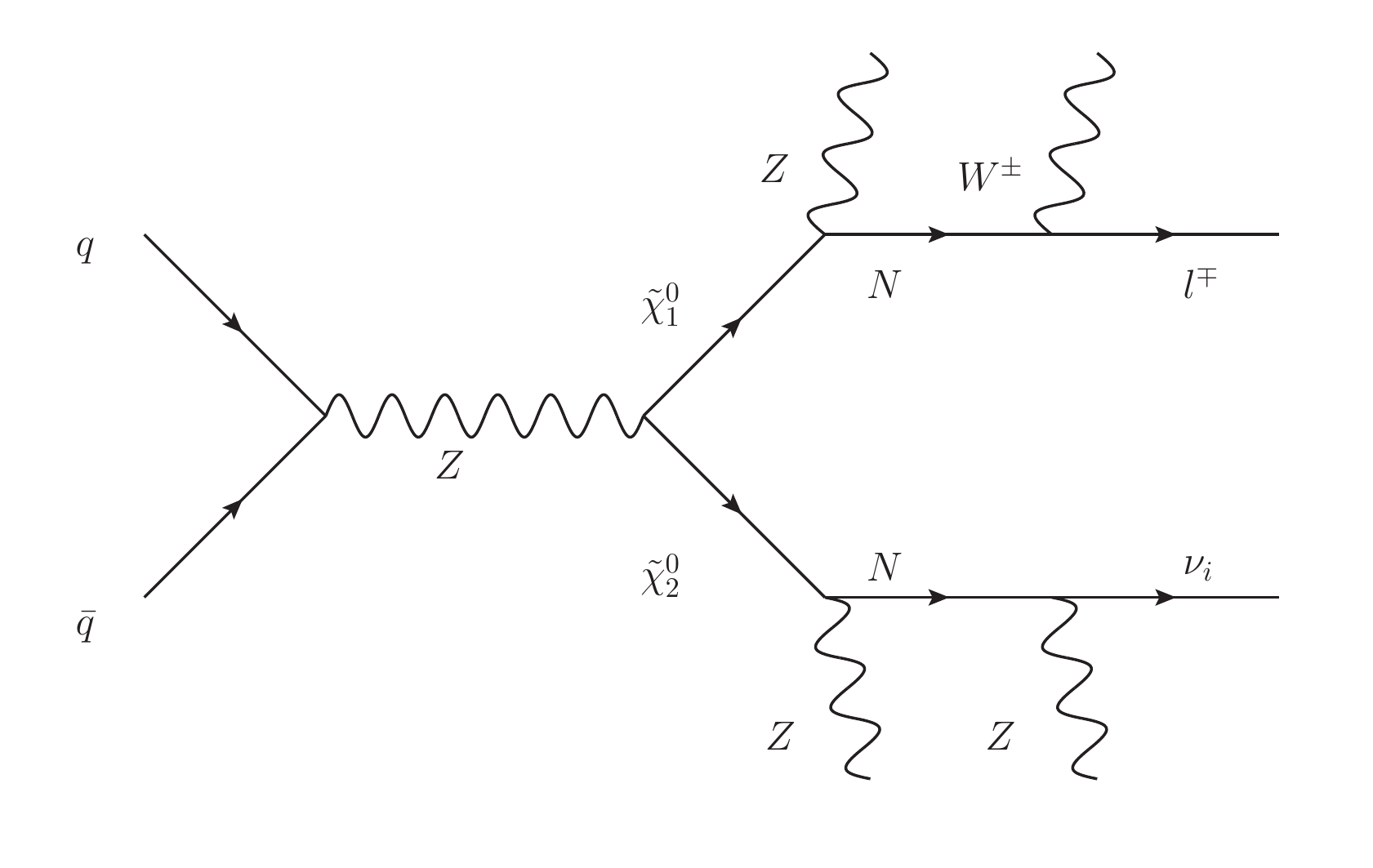}
\caption{\small Production and decay of the pseudo-Goldstone sterile neutrino
in the ``on-shell'' case ($m_N > m_Z$). In this example, two sterile neutrinos are produced
in the decay of a $\tilde \chi^0_1 \tilde \chi^0_2$ pair
($p p \to \tilde \chi^0_1 \tilde \chi^0_2 \to Z Z N N$), before decaying to $W^\pm l^\mp$
and $Z \nu_i$, respectively.}
\label{fig:production}
\end{figure}

For definiteness, we choose the following parameters
in the higgsino/electroweak gaugino sector: $\mu = 500 \GeV$,
$M_1 = 1 \TeV$, $M_2 = 2 \TeV$ and $\tan \beta = 10$. With this choice, the lightest
neutralinos $\tilde \chi^0_{1,2}$ and chargino $\tilde \chi^\pm_1$ are mostly higgsinos,
as assumed in Section~\ref{sec:model}, while $\tilde \chi^0_{3,4}$ and $\tilde \chi^\pm_2$
are gaugino-like and significantly heavier.
The mass differences between $\tilde{\chi}^0_{1}$, $\tilde{\chi}^0_{2}$ and $\tilde{\chi}^{\pm}_{1}$
are controlled by the bino mass $M_1$ and are of the order of a few GeV.
Due to their higgsino-like nature, they interact predominantly via the $W$ and $Z$ gauge bosons
and are produced with electroweak-size cross sections. By contrast, $\tilde \chi^0_{3}$, $\tilde \chi^0_{4}$
and $\tilde \chi^\pm_2$ are too heavy to be sizably produced at the LHC, and the rest
of the superpartner spectrum is assumed to be heavy enough to be decoupled from the higgsino sector.
The pseudo-Goldstone sterile neutrino, whose mass is assumed to lie in the
few $10 \GeV$ to few $100 \GeV$ range, is produced in higgsino decays
(namely, via $p p \to \tilde \chi^0_1 \tilde \chi^0_2 \to Z Z N N$,
$p p \to \tilde \chi^\pm_1 \tilde \chi^0_{1,2} \to W^\pm Z N N$
and $p p \to \tilde \chi^+_1 \chi^-_1  \to W^+ W^- N N$), 
before decaying itself through its mixing with active neutrinos.
This is illustrated in Fig.~\ref{fig:production}.
Notice that the pseudo-Goldstone sterile neutrinos are produced in pairs,
at variance with the standard scenario in which the sterile neutrino is produced through
its mixing with active neutrinos. This is due to the fact that all other higgsino decay modes
are negligible, as we will see later, and is reminiscent of $R$-parity (which is only weakly
violated in our model, $R$-parity odd couplings being suppressed by a factor
of order $\epsilon^{l - h_d}$).

While the sterile neutrino production rate and its decays are only mildly sensitive\footnote{The
active-sterile neutrino mixing
angles~(\ref{eq:V_N_alpha_general}), which together with $m_N$
control the decay rate of the sterile neutrino, are independent of these parameters.
The higgsino-sterile neutrino mixing angles~(\ref{eq:V_chi01_N}) and~(\ref{eq:V_chi02_N}),
which induce the decays $\tilde \chi^0_{1,2} \to Z N$ and $\tilde \chi^\pm_1 \to W^\pm N$,
depend (via $\lambda_0$) on $M_1$, $M_2$ and $\tan \beta$, but the branching ratios
remain close to $1$ in a broad region of the parameter space around the reference
values $M_1 = 1 \TeV$, $M_2 = 2 \TeV$ and $\tan \beta = 10$.} to the actual values
of $M_1$, $M_2$ and $\tan \beta$, they strongly depend on the $\mu$ parameter,
which controls the higgsino production cross section.
The choice $\mu = 500 \GeV$ is motivated by the negative results of the searches
for neutralinos and charginos performed at the LHC.
In Ref.~\cite{Aaboud:2017leg}, the ATLAS collaboration searched for electroweak production
of supersymmetric particles in scenarios with compressed spectra at $\sqrt{s}=13 \TeV$, with
an integrated luminosity $\mathcal{L}=36.1$ fb$^{-1}$. The constraint $\mu \gtrsim 150 \GeV$
was set from searches for mostly-higgsino neutralino/chargino pair production.
In Ref.~\cite{Sirunyan:2017lae}, the CMS collaboration searched for electroweak production
of charginos and neutralinos in multilepton final states at $\sqrt{s}=13 \TeV$, with an
integrated luminosity $\mathcal{L}=35.9$ fb$^{-1}$. Recasting the analysis done by CMS
for the channel $\tilde{\chi}^{\pm}_1 \tilde{\chi}^0_2\to W^{\pm} Z \tilde{\chi}^0_1 \tilde{\chi}^0_1$
(where $\tilde{\chi}^{\pm}_1$ and $\tilde{\chi}^0_2$ are wino-like, and $\tilde{\chi}^0_1$ is bino-like)
for the process $\tilde{\chi}^{\pm}_1 \tilde{\chi}^0_2\to W^{\pm} Z N N$
(where $\tilde{\chi}^{\pm}_1$ and $\tilde{\chi}^0_2$ are higgsino-like, and it is assumed
that the displaced vertices from sterile neutrino decays are not detected),
we obtain the lower bound $\mu \gtrsim 375 \GeV$. Taking into account the expected
improvement of this limit with the data from LHC's second run, we choose $\mu = 500 \GeV$
as a reference value.

\begin{table}[t]
\centering
\resizebox{\textwidth}{!}{
\begin{tabular}{|c|c|c|c|c|c|c|c|c|}
\hline
\rule{0mm}{5mm}
$m_{N}\, \mbox{(GeV)}$ & $|V_{Ne}|$ & $|V_{N\mu}|$ & $|V_{N\tau}|$ & $|V_{\tilde \chi^0_1 N}|$ & $|V_{\tilde \chi^0_2 N}|$
  & $|V_{\tilde \chi^0_{1,2} \alpha}|$   & $f\, \mbox{(TeV)}$ \\[0.3em]
\hline
\rule{0mm}{5mm}
 70  & $1.3\times 10^{-7}$ & $6.3\times 10^{-7}$ & $5.5\times 10^{-7}$ & $0.0062$ & $0.0069$
  & $(4.5 - 6.2) \times 10^{-6}$  &  18.9\\ [0.3em]
\hline
\rule{0mm}{5mm}
 110   & $1.0\times 10^{-7}$ & $5.0\times 10^{-7}$ & $4.4\times 10^{-7}$ & $0.0072$ & $0.0095$
  & $(4.5 - 6.2) \times 10^{-6}$   & 15.1\\[0.3em]
\hline
\end{tabular}
}
\caption{\small Values of the mixing angles between the active and sterile neutrinos ($V_{N\alpha}, \alpha = e, \mu, \tau$),
between the lightest neutralinos and the sterile neutrino ($V_{\tilde \chi^0_{1,2} N}$),
and between the lightest neutralinos and the active neutrinos ($V_{\tilde \chi^0_{1,2} \alpha}$),
computed in the maximal mixing case (formulae~(\ref{eq:V_N_alpha_max}),
(\ref{eq:V_chi01_alpha_max})--(\ref{eq:V_chi02_alpha_max})
and~(\ref{eq:V_chi01_N})--(\ref{eq:V_chi02_N})) for $m_N = 70 \GeV$ and $110 \GeV$,
respectively. Also indicated is the value of the global symmetry breaking scale $f$.
The supersymmetric parameters are chosen to be $\mu = 500 \GeV$, $M_1 = 1 \TeV$, $M_2 = 2 \TeV$
and $\tan \beta = 10$. For the neutrino parameters, we take the best fit values of Ref.~\cite{Esteban:2020cvm},
assuming normal ordering with $m_1 = 0$.}
\label{tab:example_point}
\end{table}

Having fixed $\mu$, $M_1$, $M_2$ and $\tan \beta$, and following the assumptions
made in Section~\ref{sec:model} about the neutrino sector -- namely, we neglect the
subleading contributions to the neutrino mass matrix ($\delta M_\nu = 0$) and consider
the normal mass ordering ($m_1 = 0$) -- we are left with only three real parameters:
the sterile neutrino mass $m_N$ and a complex number parametrizing the $2 \times 2$
complex orthogonal matrix $R$. Regarding the freedom associated with $R$, we
focus on the maximal mixing case (corresponding to $R = \mathbf{1}$), for which
the active-sterile mixing angles are given by Eq.~(\ref{eq:V_N_alpha_max}). For comparison,
we will also refer to the minimal mixing case (corresponding to $R_{11} = R_{22} = 0$,
$R_{12} = R_{21} = 1$), for which the active-sterile mixing angles are given by
Eq.~(\ref{eq:V_N_alpha_min}). In both cases, the CP-violating phases of the PMNS matrix
do not play a significant role and we set them to zero. For the sterile neutrino mass,
we consider two values: $m_N = 70 \GeV$ and $m_N = 110 \GeV$, corresponding
to $N$ decays via off-shell and on-shell $W$ and $Z$ gauge bosons, respectively.
The values of the mixing angles relevant for the production and decay of the sterile neutrino,
computed in the maximal mixing case, are displayed in Table~\ref{tab:example_point}
for both choices of $m_N$. The value of the global symmetry breaking scale
$f = h_d\, \mu / \lambda_0 = 15.8 \TeV / \sqrt{m_N / 100 \GeV}$ is also indicated.
We note in passing that the two example points in Table~\ref{tab:example_point}
evade the neutrinoless double beta decay constraint\footnote{This constraint follows
from the non-observation of neutrinoless double beta decay by the KamLAND-Zen experiment,
and assumes that the exchange of $N$ is the dominant contribution. We have updated
the upper limit on $V_{Ne}$ from Fig.~3 of Ref.~\cite{Faessler:2014kka}, using the lower
bound $T^{0\nu}_{1/2} (^{136}{\rm Xe}) \geq 1.07 \times 10^{26}\, \mbox{yr}$ 
from KamLAND-Zen~\cite{KamLAND-Zen:2016pfg}.}
$|V_{Ne}|\, \lesssim\, 5 \times 10^{-5}\, \sqrt{m_N / 1 \GeV}$, valid for $m_N \gtrsim 1 \GeV$~\cite{Faessler:2014kka}.

\subsection{Production and decay of the sterile neutrino}
\label{subsec:prod_decay}

Since the pseudo-Goldstone sterile neutrino is produced in decays of higgsino-like states,
its production rate is determined by the higgsino pair production cross sections
$\sigma_{\tilde{\chi}^0_1\tilde{\chi}^0_2}$, $\sigma_{\tilde{\chi}_1^{\pm}\tilde{\chi}^0_{2}}$,
$\sigma_{\tilde{\chi}_1^{\pm}\tilde{\chi}^0_{1}}$ and $\sigma_{\tilde{\chi}_1^+\tilde{\chi}_1^-}$,
and by the branching ratios for $\tilde{\chi}^0_{1}\to Z +N$,
$\tilde{\chi}^0_{2}\to Z +N$ and $\tilde{\chi}^{\pm}_1\to W^{\pm} +N$.
To compute the latter, we must consider all possible decays of $\tilde{\chi}^0_{1}$,
$\tilde{\chi}^0_{2}$ and $\tilde{\chi}^{\pm}_{1}$.
For the higgsino-like neutralinos $\tilde{\chi}^0_{1}$ and $\tilde{\chi}^0_{2}$,
the following decay modes are available:
$\tilde{\chi}^0_{1,2}\to Z +N$,
$\tilde{\chi}^0_{1,2}\to Z +\nu$, $\tilde{\chi}^0_{1,2}\to W^{\pm} +l^{\mp}$,
$\tilde{\chi}^0_{2}\to Z^{*} +\tilde{\chi}^0_1\to f\bar{f}\tilde{\chi}^0_1$,
$\tilde{\chi}^0_{2}\to W^{\pm*} +\tilde{\chi}^{\mp}_1\to f\bar{f'}\tilde{\chi}^{\mp}_1$,
$\tilde{\chi}^0_{1,2}\to N+a$, $\tilde{\chi}^0_{1,2}\to \nu+a$ and $\tilde{\chi}^0_{2}\to \tilde{\chi}^0_{1}+a$,
where $f$ and $f'$ are light fermions and $a$ is the pseudo-Nambu-Goldstone boson
associated with the spontaneous breaking of the global $U(1)$ symmetry.
For the higgsino-like chargino $\tilde{\chi}^\pm_{1}$, the possible decay modes are
$\tilde{\chi}^{\pm}_1\to W^{\pm} +N$,
$\tilde{\chi}^{\pm}_1\to W^{\pm} +\nu$, $\tilde{\chi}^{\pm}_1\to Z +l^{\pm}$,
$\tilde{\chi}^\pm_{1}\to W^{\pm*} +\tilde{\chi}^{0}_1\to f\bar{f'}\tilde{\chi}^{0}_1$
and $\tilde{\chi}^{\pm}_1\to a+l^{\pm}$.
To compute the corresponding decay rates, we generalize the standard formulae
for the couplings $Z \tilde{\chi}^0_i \tilde{\chi}^0_j$, $Z \tilde{\chi}^{\pm}_i \tilde{\chi}^{\mp}_j$
and $W^{\mp} \tilde{\chi}^{\pm}_i \tilde{\chi}^0_j$~\cite{Haber:1984rc} to include the mixing
of the active and sterile neutrinos with the neutral higgsinos and gauginos, as well as
the mixing of the charged leptons with the charged wino and higgsinos.
Then, using formulae from Ref.~\cite{Gunion:1987yh} suitably extended to our model,
we find that the decays of the higgsino-like states $\tilde{\chi}^0_{1,2}$ and $\tilde{\chi}^{\pm}_1$ 
are strongly dominated by
\begin{equation}
\tilde{\chi}^0_{1,2}\to Z +N\, , \qquad \tilde{\chi}^{\pm}_1\to W^{\pm} +N\, ,
\end{equation}
with decay rates of order  $10^{-3}$ GeV for all three processes, and branching ratios
very close to 1: $1 - {\rm BR}(\tilde{\chi}^0_{2}\to Z +N) = \mathcal{O}(10^{-5})$,
$1 - {\rm BR}(\tilde{\chi}^0_{1}\to Z +N) = \mathcal{O}(10^{-6})$
and $1 - {\rm BR}(\tilde{\chi}^{\pm}_1\to W^{\pm} +N) = \mathcal{O}(10^{-6})$.
The reason for this is the relatively large value of the mixing between the sterile neutrino
and the neutral higgsinos (represented by\footnote{As mentioned in Section~\ref{sec:model},
$V_{\tilde \chi^0_{1,2} N}$ and $V_{\tilde \chi^0_{1,2} \alpha}$, computed in the seesaw approximation,
provide good approximations for the mixing angles appearing at the
$Z \tilde \chi^0_{1,2} N / W^\pm \tilde \chi^\mp_1 N$ and $Z \tilde \chi^0_{1,2} \nu_\alpha$ vertices,
respectively.} $V_{\tilde \chi^0_{1,2} N}$ in Table~\ref{tab:example_point}),
while other decays are suppressed by smaller mixing angles.
For instance, the decays $\tilde{\chi}^0_{1,2}\to Z +\nu$ are highly suppressed by the small mixing
between neutral higgsinos and active neutrinos (represented by $V_{\tilde \chi^0_{1,2} \alpha}$
in Table~\ref{tab:example_point}), with branching ratios of order $10^{-13}$.
Similarly, the decays $\tilde{\chi}^{\pm}_1\to W^{\pm} +\nu$, $\tilde{\chi}^0_{1,2}\to W^{\pm} +l^{\mp}$
and $\tilde{\chi}^{\pm}_1\to Z +l^{\pm}$ are suppressed by the small charged lepton--charged higgsino
and active neutrino--neutral higgsino mixings.
As for the 3-body decays $\tilde{\chi}^0_{2}\to Z^{*} +\tilde{\chi}^0_1\to f\bar{f}\tilde{\chi}^0_1$,
$\tilde{\chi}^0_{2}\to W^{\pm*} +\tilde{\chi}^{\mp}_1\to f\bar{f'}\tilde{\chi}^{\mp}_1$
and $\tilde{\chi}^\pm_{1}\to W^{\pm*} +\tilde{\chi}^{0}_1\to f\bar{f'}\tilde{\chi}^{0}_1$,
they are suppressed by phase-space kinematics, with branching ratios of order $10^{-8}$.
Finally, the decays involving the PNGB $a$, induced by its coupling to the down-type higgsino
(see Eq.~(\ref{eq:aff_tree}) in Appendix~\ref{app:PNGB}), are suppressed by the global
symmetry breaking scale $f$. If kinematically allowed and not phase-space suppressed,
$\tilde{\chi}^0_{2}\to \tilde{\chi}^0_{1} + a$ has a branching ratio of order $10^{-6}$,
larger than other decays but still well below the dominant $\tilde \chi^0_2$ decay mode,
$\tilde{\chi}^0_{2}\to Z +N$. The decay modes $\tilde{\chi}^0_{1,2}\to N+a$,
$\tilde{\chi}^0_{1,2}\to \nu+a$ and $\tilde{\chi}^{\pm}_1\to a+l^{\pm}$ are always
kinematically allowed, but are suppressed by small mixing angles, in addition to the
$1/f$ suppression.

We can therefore neglect all decays of the higgsino-like states but the ones
with a sterile neutrino in the final state, $\tilde{\chi}^0_{1,2}\to Z +N$ and $\tilde{\chi}^{\pm}_1\to W^{\pm} +N$.
We checked that these decays are prompt and do not lead to displaced vertices.
The sterile neutrinos are therefore produced in pairs, with a cross section of electroweak size
(namely, the higgsino pair production cross section).
They subsequently decay via on-shell or off-shell W and Z bosons, as shown in Fig.~\ref{fig:production}.
Since these decays involve the mixing angles of the sterile neutrino with the active ones,
which are of order $10^{-7} - 10^{-6}$ (see Table~\ref{tab:example_point}), they may
lead to observable displaced vertices, depending on the sterile neutrino mass $m_N$.

\begin{figure}[t]
 \centering
 \includegraphics[height=7cm]{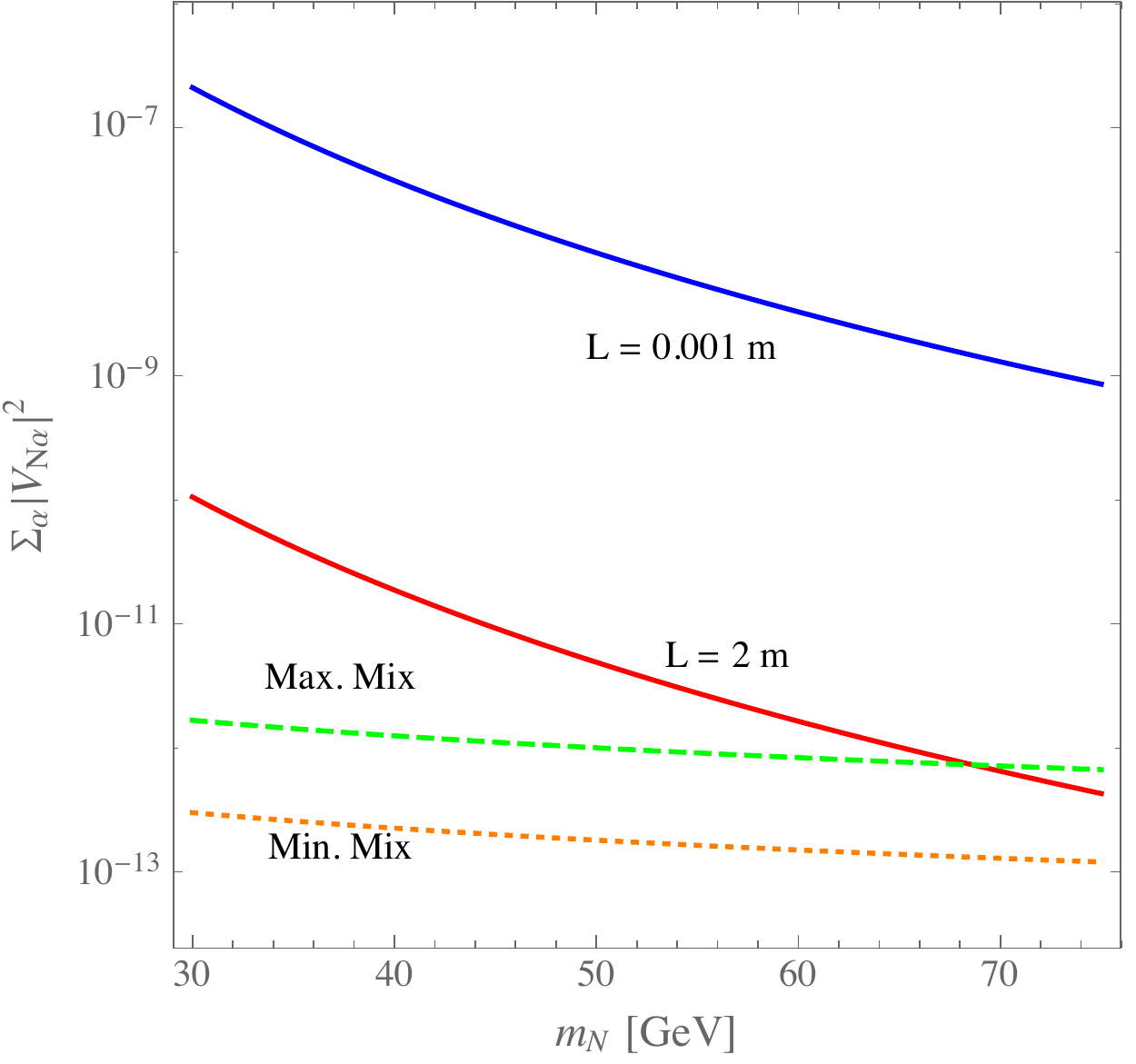}
  \includegraphics[height=7cm]{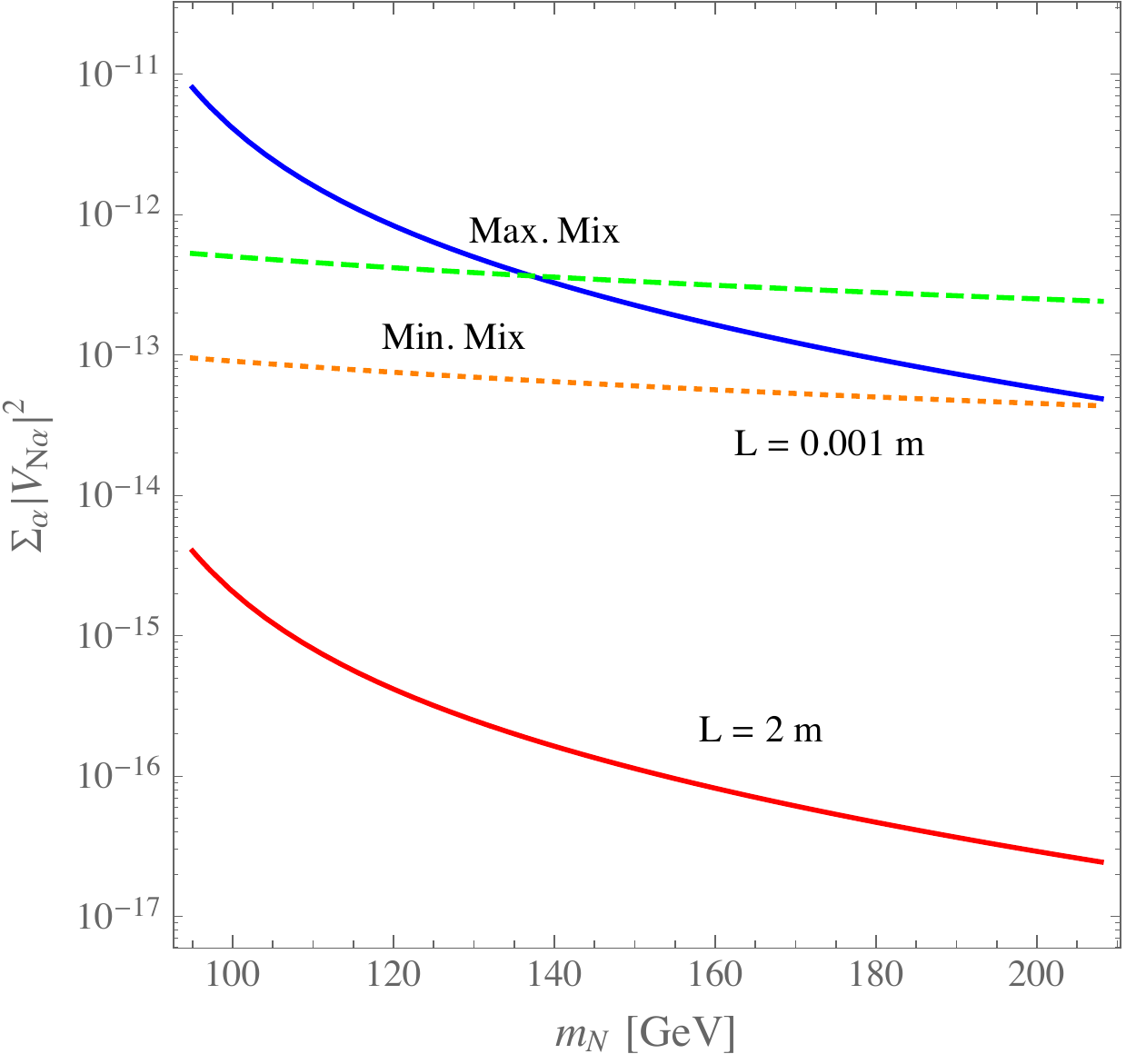}
\caption{\small Sterile neutrino decay length $L$ as a function of its mass ($m_{N}$) and of the
active-sterile neutrino mixing ($\sum_\alpha |V_{N \alpha}|^2$) at the LHC with $\sqrt s = 14 \TeV$,
assuming the same model parameters as in Table~\ref{tab:example_point}.
The blue and red solid lines correspond to $L = 1\, \mbox{mm}$ and $L = 2\, \mbox{m}$, respectively.
The green long-dashed line (resp. the orange short-dashed line) is the prediction of the model
for $\sum_{\alpha} |V_{N \alpha}|^2$ in the maximal mixing case (resp. in the minimal mixing case).
The predicted decay length falls in the range $[1\, \mbox{mm}, 2\, \mbox{m}]$ when the
relevant dashed line lies between the blue and red solid lines.}
 \label{fig:mixing}
\end{figure}

To identify the range of $m_N$ values for which this is the case, we calculated
the sterile neutrino decay length in the (numerically very good)
approximation where the $W N \ell_\alpha$ and $Z N \nu_\alpha$ couplings
are expressed in terms of the mixing angles $V_{N \alpha}$.
For the ``off-shell case'' $m_N < m_W$, where $N$ decays via off-shell $W$ and $Z$
gauge bosons, we used the formulae provided in Ref.~\cite{Helo:2010cw}
for the sterile neutrino partial decay widths.
In the ``on-shell case'' $m_N > m_Z$, one can derive a simple formula for the sterile neutrino
decay length $L$ by neglecting the masses of the final state leptons:
\begin{equation}
  L\ \simeq\ \beta\gamma\, \frac{5.99\times 10^{-16}\; {\rm m}}{\left( \frac{m_N}{100 \GeV} \right)^3
    \left[\left(1-\frac{m^2_Z}{m^2_N}\right)^2\left(1+2\frac{m^2_Z}{m^2_N}\right)
    +2\left(1-\frac{m^2_W}{m^2_N}\right)^2\left(1+2\frac{m^2_W}{m^2_N}\right)\right]\sum_{\alpha}|V_{N\alpha}|^2}\ .
\end{equation}
Fig.~\ref{fig:mixing} shows the region of the ($m_N$, $\sum_{\alpha} |V_{N \alpha}|^2$)
parameter space where $1\, \mbox{mm} \leq L \leq 2\, \mbox{m}$,
in the off-shell (left panel) and on-shell (right panel) cases. Also shown is the prediction
of the model for $\sum_{\alpha} |V_{N \alpha}|^2$ as a function of $m_N$,
in the minimal mixing and maximal mixing cases.
In theses plots, the $\beta \gamma$ of the sterile neutrino is approximated by the
$(\beta \gamma)_{\rm eff}$ introduced in Section~\ref{sec:motivation}.
The value of $(\beta \gamma)_{\rm eff}$ was estimated by simulating the pair production
of $500 \GeV$ higgsino-like neutralinos at the 14 TeV LHC with {\tt MadGraph\_aMC@NLO 2.6}~\cite{Alwall:2014hca},
taking the peak value of their $\beta\gamma$ distribution and using it to compute the $\beta\gamma$
of the sterile neutrino, assuming it is emitted in the same direction as the parent neutralino.
We checked that the value of $(\beta \gamma)_{\rm eff}$ does not change much when
one considers slightly larger values of $\mu$.

The area delineated by the blue and red solid lines in Fig.~\ref{fig:mixing}
gives an idea of the range of sterile neutrino masses and mixing angles
that can lead to observable displaced vertices at the $14 \TeV$ LHC
(for definiteness, and without entering the characteristics of the ATLAS and CMS detectors,
we take $1\, \mbox{mm}$ as the minimal displacement detectable by the tracking system,
and $2\, \mbox{m}$ as the distance above which the sensitivity to displaced vertices drops).
Comparing these curves with the predictions of the model in two benchmark cases,
minimal mixing and maximal mixing, one can see that a pseudo-Goldstone sterile
neutrino with a mass of order $100 \GeV$ (from $60 \GeV$ or less
in the maximal mixing case to about $200 \GeV$ for minimal mixing) is accessible
to displaced vertex searches at the LHC.
For illustration, we give in Table~\ref{tab:branchingratios}, for two representative values
of the sterile neutrino mass
and in the maximal mixing case, the fraction of decays occuring
between $1\, \mbox{mm}$ and $2\, \mbox{m}$ from the collision point, as well as the percentages
of final states $l + 2\, {\rm jets}$ ($l = e, \mu, \tau$) and $\nu + 2\, {\rm jets}$.

It is interesting to note that the values of the mixing angles that can be probed
in our model are much smaller than in the standard scenario, where the heavy sterile
neutrino only mixes with the active neutrinos (see the discussion at the beginning
of Section~\ref{sec:motivation}).
This is due to the fact that, in the pseudo-Goldstone sterile neutrino scenario,
the production cross section is of electroweak size, while it is suppressed by the $|V_{N \alpha}|^2$
in the standard case, thus limiting the sensitivity to these parameters.
As a result, the region of the parameter space that can be probed via displaced vertex searches at the LHC
shifts from roughly $m_N \leq 40 \GeV$, $\sum_{\alpha} |V_{N \alpha}|^2 \geq {\rm few}\, 10^{-9}$
in the standard scenario~\cite{Drewes:2019fou} to $m_N \approx ({\rm few}\, 10 \GeV - 200 \GeV)$,
$\sum_{\alpha} |V_{N \alpha}|^2 \approx ({\rm few}\, 10^{-14} - 10^{-12})$ in our model.
This rather narrow range of values for the $V_{N\alpha}$ is a peculiarity of the model,
which predicts some correlation between the mixing angles and the sterile neutrino mass.
Other models belonging to the same class (i.e., in which the active-sterile mixing
does not enter the production of the sterile neutrino, but is responsible for its decays)
may cover a larger part of the parameter space consistent with observable displaced
vertices at the LHC (area between the blue and red solid lines in Fig.~\ref{fig:mixing}).

\begin{table}[t]
\centering
\begin{tabular}{|c||c|c|}
\hline
\rule{0mm}{5mm}
          &  $m_N=70$ GeV  & $m_N=110$ GeV \\[0.3em]
\hline
\rule{0mm}{5mm}
 $(\beta \gamma)_{\rm eff}$ & 6.4 & 4 \\ [0.3em]
\hline
\rule{0mm}{5mm}
 Decay length & 1.78 m & 3.49 mm \\ [0.3em]
 \hline
 \rule{0mm}{5mm}
 Fraction of N decays in $[1\, \mbox{mm}, 2\, \mbox{m}]$ & 67.4 $\%$  & 75.1 $\%$ \\ [0.3em]
 \hline
\rule{0mm}{5mm}
 e + 2 jets &  0.71 $\%$ & 0.90 $\%$  \\ [0.3em]
 \hline
\rule{0mm}{5mm}
$\mu$ + 2 jets & 17.5 $\%$ & 22.0 $\%$  \\ [0.3em]
 \hline
\rule{0mm}{5mm}
$\tau$ + 2 jets & 13.5 $\%$ & 17.0 $\%$  \\ [0.3em]
\hline
\rule{0mm}{5mm}
$\nu$ + 2 jets & 13.0 $\%$ & 10.8 $\%$  \\ [0.3em]
\hline
\end{tabular}
\caption{\small Percentage of sterile neutrinos decaying to the final states $l + 2\, {\rm jets}$ ($l = e, \mu, \tau$)
and $\nu + 2\, {\rm jets}$ between $1\, \mbox{mm}$ and $2\, \mbox{m}$ from the production point,
for $m_N = 70 \GeV$ and $110 \GeV$ and in the maximal mixing case.
The other model parameters are chosen as in Table~\ref{tab:example_point}.
Also shown are the value of the $(\beta \gamma)_{\rm eff}$ parameter, estimated as explained
in the text, the sterile neutrino decay length and the total fraction of decays occurring between 
$1\, \mbox{mm}$ and $2\, \mbox{m}$.
}
\label{tab:branchingratios}
\end{table}
%

\subsection{Reconstruction of the active-sterile neutrino mixing angles}
\label{subsec:reconstruction}

Let us now study more quantitatively the signals arising from the production and decay
of the pseudo-Goldstone sterile neutrino, and outline a strategy for
measuring the active-sterile mixing angles $V_{N \alpha}$.
As discussed before, the sterile neutrino is produced in the decays of higgsino-like states,
together with a $W$ or $Z$ boson, with a branching ratio close to $100 \%$.
Given the choice $\mu = 500 \GeV$ and the values of $m_N$ considered,
the decay products of these gauge bosons are boosted, providing triggers in the form
of high $p_T$ leptons for the signal we want to analyse,
namely the displaced vertices from the decays of the mostly sterile states $N$.
Provided that the decay products of $N$ can be reconstructed, one can determine
its total decay width as well as its partial decay widths to different final states,
from which the active-sterile mixing angles $V_{N \alpha}$ can be extracted.

The sterile neutrino pair production rate is determined by the production cross sections
for the pairs of higgsino-like states $\tilde{\chi}^0_1\tilde{\chi}^0_2$, $\tilde{\chi}_1^{\pm}\tilde{\chi}^0_{1}$,
$\tilde{\chi}_1^{\pm}\tilde{\chi}^0_{2}$ and $\tilde{\chi}_1^+ \tilde{\chi}_1^-$.
These cross sections have been computed for the $13 \TeV$ LHC at NLO-NLL with MSTW2008nlo90cl
PDFs~\cite{Fuks:2012qx,Fuks:2013vua}, and the channel $pp \to \tilde{\chi}_1^{\pm}\tilde{\chi}^0_{2}$
has also been calculated at $\sqrt{s}=14 \TeV$. Using the ratio of the $\tilde{\chi}_1^{\pm}\tilde{\chi}^0_{2}$
production cross sections at $\sqrt{s}=13 \TeV$ and $\sqrt{s}=14 \TeV$ to rescale the other channels
from $\sqrt{s} = 13 \TeV$ to $\sqrt{s} = 14 \TeV$, we obtain
$\sigma_{\tilde{\chi}_1^{\pm}\tilde{\chi}^0_{1}} \simeq \sigma_{\tilde{\chi}_1^{\pm}\tilde{\chi}^0_{2}} \simeq (13.1 \pm 0.8)\, {\rm fb}$,
$\sigma_{\tilde{\chi}_1^+ \tilde{\chi}_1^-} \simeq (7.6 \pm 0.4)\, {\rm fb}$,
and $\sigma_{\tilde{\chi}^0_1\tilde{\chi}^0_2} \simeq (7.0 \pm 0.3)\, {\rm fb}$
for $m_{\tilde{\chi}_1^{0}} \simeq m_{\tilde{\chi}_2^{0}} \simeq m_{\tilde{\chi}_1^{\pm}} \simeq 500 \GeV$,
where the errors take into account the scale and parton distribution function uncertainties.
Since ${\rm BR} (\tilde{\chi}^{\pm}_1 \to W^{\pm} +N) = {\rm BR} (\tilde{\chi}^{0}_{1,2} \to Z +N) = 1$
to an excellent approximation, $\sigma_{\tilde{\chi}_1^{\pm}\tilde{\chi}^0_{1,2}} \equiv
\sigma_{\tilde{\chi}_1^{\pm}\tilde{\chi}^0_{1}} + \sigma_{\tilde{\chi}_1^{\pm}\tilde{\chi}^0_{2}}$,
$\sigma_{\tilde{\chi}_1^+ \tilde{\chi}_1^-}$ and $\sigma_{\tilde{\chi}^0_1\tilde{\chi}^0_2}$
can be identified with the production cross sections for $W^{\pm} Z NN$, $W^+ W^- N N$
and $Z Z N N$, respectively. Even though it is possible to distinguish experimentally
between the different production channels\footnote{One can distinguish between
the different production channels by focusing on the leptonic decays of the $W$ and $Z$ bosons:
$Z Z \to l^{+} l^{-} l'^+ l'^-$, $Z W^{\pm} \to l^{+} l^{-} l'^{\pm} \nu$
and $W^+ W^- \to l^{+} \nu\, l'^{-} \bar \nu$.},
we shall only consider the total sterile neutrino production cross section
$\sigma_0 \equiv \sigma_{\tilde{\chi}_{1}^{\pm}\tilde{\chi}^0_{1,2}}
+ \sigma_{\tilde{\chi}_1^+ \tilde{\chi}_1^-} + \sigma_{\tilde{\chi}^0_1\tilde{\chi}^0_2}$
in the following, using $\sigma_0 = 40.8\, {\rm fb}$ at $\sqrt{s}=14 \TeV$.

Once produced, the sterile neutrinos decay via on-shell or off-shell W and Z bosons.
Decays that proceed
via a $W$ boson produce a charged lepton, whose flavour $\alpha$ can in principle be identified.
The corresponding branching ratio is therefore proportional to $|V_{N \alpha}|^2$.
Instead, in decays mediated by a $Z$ boson, the three neutrino flavours are indistinguishable
and the branching ratios are proportional to $\sum_\alpha |V_{N \alpha}|^2$.
To sketch the procedure for determining the active-sterile mixing angles from experimental data,
let us have a look at Eq.~(\ref{eq:number_decay}), which provides a simplified expression
(in the approximation where all sterile neutrinos have the same $\beta \gamma$)
for the number of events $N_i$ corresponding to a given final state $i$.
In this formula, the sterile neutrino decay width $\Gamma$
(which, as explained in Section~\ref{sec:motivation}, can be extracted from the shape
of the distribution of displaced vertices) and $(\beta \gamma)_{\rm eff}$ (or, in a more
proper treatment, the $\beta \gamma$ distribution of the sterile neutrinos, which can
be reconstructed experimentally) are assumed to be known.
The measurement of $N_i$ thus provides us with $N_0 \Gamma_i$, where $N_0$ is the number
of sterile neutrino produced and $\Gamma_i$ their partial decay width into the final state $i$.
Using the theoretical expression
for $\Gamma_i$, we straightforwardly convert $N_0 \Gamma_i$ into $N_0 |V_{N \alpha}|^2$
(where $\alpha$ is the relevant lepton flavour) if the decay proceeds via a $W$ boson,
or into $N_0 \sum_\alpha |V_{N \alpha}|^2$ if it is mediated by a $Z$ boson.
By considering different final states, we can in principle determine all active-sterile mixing angles
$V_{N \alpha}$ and break the degeneracy with $N_0$.

In practice, and taking into account the fact that the sterile neutrinos are produced in pairs,
we will focus on final states with two displaced vertices involving either two charged leptons,
jets and no missing transverse energy (MET), or one charged lepton, jets and MET.
The first category of events ($l + l' + {\rm jets}$) can be unambiguously assigned to both $N$'s
decaying via a $W$ boson into a charged lepton and two jets, whereas the second category
($l + \nu + {\rm jets}$) can be unambiguously assigned to one $N$ decaying as
$N \to  l^{\pm} W^{\mp (*)} \to l^{\pm} + 2\, {\rm jets}$ and the other one as
$N \to \nu Z^{(*)} \to \nu + 2\, {\rm jets}$.
We assume that the charged leptons
and jets can be properly assigned to one of the two displaced vertices.
This can be accomplished for example by demanding that the two charged leptons
in the first category of events have different flavours.

Restricting to the case where the charged leptons are electrons or muons (which are
easily identified at the LHC), we are left with five different final states, with corresponding
number of events $N_{ll'}$ and $N_{l\nu}$ ($l, l' =e,\mu$):
\begin{eqnarray}
N_{ee} &\propto &  N_{0} |V_{N e}|^4\, ,  \label{eq:Nee} \\
N_{\mu\mu} &\propto &  N_{0} |V_{N \mu}|^4\, , \\
N_{e\mu} &\propto &  N_{0} |V_{N e}|^2 |V_{N \mu}|^2\, , \\
N_{e\nu} &\propto &   N_{0} |V_{N e}|^2(|V_{N e}|^2 +|V_{N \mu}|^2+ |V_{N \tau}|^2)\, ,\\
N_{\mu\nu} &\propto &   N_{0} |V_{N \mu}|^2(|V_{N e}|^2 +|V_{N \mu}|^2+ |V_{N \tau}|^2)\, ,
\label{eq:Nmunu}
\end{eqnarray}
where $N_0 = \mathcal{L}\, \sigma_0$, with $\sigma_0$ the sterile neutrino pair production
cross section and $\mathcal{L}$ the integrated luminosity of interest.
Since the proportionality factors in Eqs.~(\ref{eq:Nee})--(\ref{eq:Nmunu})
are known (they depend on $\Gamma$, $m_{N}$, $\beta\gamma$,
all of which can be determined from the experimental data,
and on the masses of the final state particles), we can solve these equations for 
$N_{0}^{1/2} |V_{N e}|^2$, $N_{0}^{1/2} |V_{N \mu}|^2$ and $N_{0}^{1/2} |V_{N \tau}|^2$
in terms of three suitably chosen numbers of events, for instance $N_{\mu\mu}$,
$N_{e\mu}$ and $N_{\mu\nu}$.
Using this experimental input alongside the theoretical expressions for the partial decay widths
given in Ref.~\cite{Helo:2010cw}
for the off-sell case, and in Ref.~\cite{Gunion:1987yh} for the on-shell case
(with the modifications of the couplings needed to adapt the formulae to our model),
we obtain $N^{1/2}_0 \Gamma$.
Since $\Gamma$ can be reconstructed from the distribution of displaced vertices,
we are finally able to break the degeneracy between $N_0$ and the active-sterile neutrino
mixing angles $|V_{N \alpha}|$, and solve for the production cross section $\sigma_0 = N_0 / \mathcal{L}$.

\begin{table}[t]
\centering
\begin{tabular}{|c||c|c|}
\hline
\rule{0mm}{5mm}
 Process  &  $m_N=70$ GeV  & $m_N=110$ GeV \\[0.3em]
\hline
\rule{0mm}{5mm}
 $\sigma_{ee}$ & 0.002 fb& 0.003 fb \\ [0.3em]
 \hline
\rule{0mm}{5mm}
 $\sigma_{e\mu}$ & 0.10 fb & 0.16 fb \\ [0.3em]
 \hline
\rule{0mm}{5mm}
 $\sigma_{\mu\mu}$ & 1.25 fb & 1.98 fb \\ [0.3em]
\hline
\rule{0mm}{5mm}
$\sigma_{e \nu}$ & 0.076 fb & 0.079 fb \\ [0.3em]
\hline
\rule{0mm}{5mm}
$\sigma_{\mu \nu}$ & 1.86 fb & 1.94 fb \\ [0.3em]
\hline
\end{tabular}
\caption{\small Cross sections corresponding to the final states $l + l' + {\rm jets}$ and $l + \nu + {\rm jets}$
($l, l' = e, \mu$) at the $14 \TeV$ LHC (using $\sigma_0 = 40.8\, {\rm fb}$ and omitting uncertainties),
for $m_N = 70 \GeV$ and $m_N = 110 \GeV$ and in the maximal mixing case.
The other model parameters are the same as in Table~\ref{tab:example_point}.}
\label{tab:finalcrosssections}
\end{table}

We present in Table~\ref{tab:finalcrosssections} the predictions of our model
for the cross sections corresponding to the final states considered above,
assuming the same parameters as before. These cross sections are obtained
by multiplying the sterile neutrino pair production cross section by the branching
ratios for the relevant decay channels, weighted by the fraction of decays occuring
between $1\, \mbox{mm}$ and $2\, \mbox{m}$ (see Table~\ref{tab:branchingratios}).
The expected number of events $N_{ee}$, $N_{e\mu}$, $N_{\mu\mu}$, $N_{e\nu}$
and $N_{\mu\nu}$ (before cuts and efficiencies) can be obtained by multiplying
these cross sections by the relevant integrated luminosity.
Apart from $\sigma_{ee}$, these cross sections are large enough to be able
to be probed during the run 3 of the LHC.
With an expected integrated luminosity of $3\, \mbox{ab}^{-1}$, the HL-LHC
would be able to probe a larger portion of the model parameter space, corresponding
to a broader range of sterile neutrino masses.

To conclude this section, let us recall that the above results were obtained
assuming that the orthogonal matrix $R$ in Eq.~(\ref{eq:V_N_alpha_general}) is real.
Relaxing this assumption would allow for larger values of the active-sterile neutrino
mixing angles $V_{N \alpha}$
and would therefore enlarge the region of the $(m_N, V_{N \alpha})$ parameter
space that can be probed by displaced vertex searches.
It is interesting to note, however, that the $V_{N \alpha}$ predicted in the real
case correspond to typical, ``natural'' values for these parameters.
To see this, let us rewrite the active neutrino mass matrix as
\begin{equation}
  (M_\nu)_{\alpha \beta}\, \simeq\, -\, m_N V_{N \alpha} V_{N \beta}
    - m_{\rm eff} V_{\tilde \chi^0_1 \alpha} V_{\tilde \chi^0_1 \beta}\, ,
\label{eq:M_nu}
\end{equation}
where $V_{\tilde \chi^0_1 \alpha}$ is the mixing between the lightest neutralino
and the active neutrino of flavour $\alpha$, and
$m_{\rm eff} \equiv 2 (c^2_W M_1 + s^2_W M_2) m^2_Z \cos^2 \beta / (M_1 M_2)
\simeq 0.1 \GeV (2 \TeV / M_2) (10 / \tan \beta)^2$ (for $M_2 = 2 M_1$).
In the absence of cancellations between the two terms in Eq.~(\ref{eq:M_nu}),
neutrino data requires
\begin{equation}
  |V_{N \alpha}|\, \lesssim\, (1-5) \times 10^{-7}\, \sqrt{\frac{100 \GeV}{m_N}}\ , \qquad
  |V_{\tilde \chi^0_1 \alpha}|\, \lesssim\, (3-20) \times 10^{-6}\, \sqrt{\frac{100 \GeV}{m_N}}\, ,
\end{equation}
where at least one of the two inequalities should be saturated.
These numbers are in agreement with the ones displayed in Table~\ref{tab:example_point}
(which corresponds to the maximal mixing case, i.e. $R = \mathbf{1}$). Much larger
values of the active-sterile neutrino mixing angles, which can be obtained in the
case of a complex $R$ matrix, would imply that the observed neutrino mass
scale arises from a cancellation between two unrelated contributions.

\section{Conclusions}    %
\label{sec:conclusion}    %

Low-scale models of neutrino mass generation often feature sterile neutrinos
with masses in the GeV-TeV range, which can be produced at colliders through
their mixing with the Standard Model neutrinos.
In this paper, we have considered an alternative scenario in which the sterile neutrino
is produced in the decay of a heavier particle, such that its production rate can
be sizable even if the active-sterile neutrino mixing angles are small.
As we have shown, these mixing angles can be determined from the decays of the sterile neutrino,
provided that they lead to observable displaced vertices and that different categories
of final states can be reconstructed experimentally.
Since the sterile neutrino production cross section is not suppressed by the active-sterile mixing,
displaced vertex searches can probe very small values of the $V_{N \alpha}$
-- as small as the ones predicted by the na\"{i}ve seesaw formula
$V_{N\alpha} \sim \sqrt{m_\nu/m_N}\,$, or even smaller.

We presented an explicit realization of this scenario in which the sterile neutrino
is the supersymmetric partner of the pseudo-Nambu-Goldstone boson of a spontaneously
broken global $U(1)$ symmetry. The pseudo-Goldstone sterile neutrino gets its mass from
supersymmetry breaking and mixes with the active neutrinos and the neutralinos as a
consequence of the global symmetry. Assuming relatively heavy gauginos, the sterile
neutrino is produced in the decays of higgsino-like states and decays subsequently
via its mixing with active neutrinos. Once the Standard Model neutrino parameters
are fixed to their measured values, the active-sterile neutrino mixing angles are predicted
in terms of the sterile neutrino mass and of a complex orthogonal matrix $R$.
Assuming that this matrix is real,
we have shown that a sterile neutrino with a mass between a few $10 \GeV$ and $200 \GeV$
can be observed at the LHC, and outlined a strategy for reconstructing experimentally
the active-sterile neutrino mixing angles (which in this mass interval range from
$10^{-7}$ to $10^{-6}$).
Relaxing the assumption that $R$ is real would have the effect of allowing for larger mixing angles,
and would therefore enlarge the region of the $(m_N, V_{N \alpha})$ parameter
space that can be probed by displaced vertex searches.


\paragraph*{Acknowledgments}

We are grateful to Marc Besan\c{c}on, Federico Ferri and Gautier Hamel de Monchenault
for useful discussions about heavy neutral lepton searches at the LHC, and to Tony Gherghetta
for an interesting discussion about the model of Section~\ref{sec:model}.
This work has been supported in part by the European Research Council (ERC) Advanced Grant Higgs@LHC, 
by the European Union Horizon 2020 Research and Innovation Programme under the Marie Sklodowska-Curie
grant agreements No. 690575 and  No. 674896, and by CONICET and ANPCyT under projects PICT 2016-0164 
and PICT 2017-0802.
S.~L. acknowledges the hospitality of the University of Washington and of the Fermilab
Theoretical Physics Department while working on this project.
A.~M. warmly thanks IPhT Saclay for its hospitality during the completion of this work.


\appendix

\section{Supersymmetry breaking and $R$-parity violation}  %
\label{app:alignment}                                                              %

\renewcommand{\theequation}{A.\arabic{equation}}
\setcounter{equation}{0}  

In this Appendix, we derive the expressions~(\ref{eq:parameters}) for the superpotential
parameters $\mu_0$, $\mu_i$, $\lambda_0$ and $\lambda_i$,
and show that the ones that violate $R$-parity are suppressed by $\xi \sim \epsilon^{l - h_d}$,
taking into account the interplay between $R$-parity violation and supersymmetry breaking.
In the absence of $R$-parity conservation, the scalar potential contains $R$-parity violating
soft terms which induce small vevs for the sneutrinos,
$v_i \equiv \langle \tilde \nu_i \rangle \ll v_0 \equiv \langle h^0_d \rangle$ (as we will see later,
the hierarchy $v_i \ll v_0$ follows from the $U(1)$ symmetry). It is convenient to redefine
the superfields $H_d$ and $L_i$ in such a way that the vev is carried solely by the scalar
component of $H_d$, and the $L_i$'s can be identified with the physical lepton doublet superfields.
This can be done by the following $SU(4)$ rotation on the 4-vector $(H_d, L_i)$:
\begin{equation}
  H_d \, \to\,  \frac{v_0}{v_d}\, H_d - \sum_i \frac{v^*_i}{v_d}\, L_i + \dots\, ,\qquad
  L_i \,  \to \, \frac{v^*_0}{v_d}\, L_i + \frac{v_i}{v_d}\, H_d +\, \dots\, ,
\label{eq:SU4_rotation}
\end{equation}
where $v_d \equiv \sqrt{|v_0|^2 + \sum_i |v_i|^2}\,$, and the dots stand for corrections
of order $(v_i/v_d)^2$. Finally, we diagonalize the charged lepton Yukawa couplings
by means of the unitary transformations $L_i \to \sum_j R^e_{ji} L_j$
and $\bar e_i \to \sum_j R^{\bar e}_{ji} \bar e_j$, and similarly for the down-type quarks:
$Q_i \to \sum_j R^d_{ji} Q_j$, $\bar d_i \to \sum_j R^{\bar d}_{ji} \bar d_j$.
After these field redefinitions, the superpotential keeps the same form as in
Eq.~(\ref{eq:W_Phi_hat}), but with diagonal Yukawa couplings and modified parameters
(at leading order in $v_i/v_d$):
\begin{equation}
  \mu_0 = \hat{\vec{\mu}} \cdot \frac{\vec{v}}{v_d}\, ,  \qquad
  \mu_i = \sum_j R^e_{ij} \left( \frac{v^*_0}{v_d}\, \hat \mu_j - \frac{v^*_j}{v_d}\, \hat \mu_0 \right) ,
\label{eq:mu_i}
\end{equation}
\begin{equation}
  \lambda_0 = h_d\, \frac{\mu_0}{f} + (l - h_d) \sum_i \frac{v_i}{v_d}\, \frac{\hat \mu_i}{f}\, \simeq\, h_d\, \frac{\mu_0}{f}\, ,
  \qquad \lambda_i = l\, \frac{\mu_i}{f} + (l - h_d) \sum_j R^e_{ij}\, \frac{v^*_j}{v_d}\, \frac{\hat \mu_0}{f}\, ,
\label{eq:lambda_i}
\end{equation}
where $\mu \equiv \sqrt{|\hat \mu_0|^2 + \sum_i |\hat \mu_i|^2} =\! \sqrt{|\mu_0|^2 + \sum_i |\mu_i|^2}\, $.
Similarly, the Yukawa couplings
and the trilinear $R$-parity violating couplings $\lambda_{ijk}$, $\lambda'_{ijk}$
can be written in terms of the original superpotential parameters of Eq.~(\ref{eq:W_Phi_hat}),
but the expressions are not illuminating and we do not write them.
Due to the hierarchy among the original parameters (namely,
$\hat \lambda_{ijk} \ll \hat \lambda^e_{jk}$ and $\hat \lambda'_{ijk} \ll \hat \lambda^d_{jk}$)
and between the vevs ($v_i \ll v_0$), $\lambda_{ijk}$ and $\lambda'_{ijk}$
are suppressed relative to the charged lepton and down quark Yukawa couplings,
and the experimental constraints on $R$-parity violation are easily satisfied.

Following Refs.~\cite{Banks:1995by,Binetruy:1997sm}, we introduce the angle $\theta$ measuring the
misalignment\footnote{With the definition~(\ref{eq:alignment_angle}), $\theta$ vanishes when
$\vec{v} = z \hat{\vec{\mu}}^*$ ($z \in \mathbb{C}^*$), which is referred to as alignment condition.
The more $\hat{\vec{\mu}}$ and $\vec{v}$ depart from this condition, the larger the misalignment angle $\theta$
(or the parameter $\xi$ defined below). Conversely, $\cos \theta \simeq 1$ (or equivalently $\xi \ll 1$)
when $\hat{\vec{\mu}}$ and $\vec{v}$ are approximately aligned.}
between the 4-vectors $\hat{\vec{\mu}} = (\hat \mu_0, \hat \mu_1, \hat \mu_2, \hat \mu_3)$
and $\vec{v} = (v_0, v_1, v_2, v_3)$:
\begin{equation}
  \cos \theta\, \equiv\, \left| \frac{\hat{\vec{\mu}} \cdot \vec{v}}{\mu v_d} \right|\, .
\label{eq:alignment_angle}
\end{equation}
Due to $v_i \ll v_0$ and, from Eq.~(\ref{eq:hierarchy_initial_parameters}), $\hat \mu_i \ll \hat \mu_0$,
we have $\cos \theta \simeq 1$ and
\begin{equation}
  1 - \cos^2 \theta\, \simeq\, \sum_i\, |\xi_i|^2\, ,  \qquad
    \xi_i\, \equiv\, \frac{\hat \mu_i v^*_0 - \hat \mu_0 v^*_i}{\mu v_d}\ \ll\, 1\ .
\label{eq:xi}
\end{equation}
Comparing Eq.~(\ref{eq:mu_i}) with Eq.~(\ref{eq:xi}), one can see that $|\mu_0| = \mu \cos \theta \simeq \mu$
and $\mu_i = \mu \sum_j R^e_{ij}\, \xi_j\, $: the size of the bilinear $R$-parity violating parameters $\mu_i$
is controlled by the small misalignment variables $\xi_i$.
To make the relative size of the superpotential parameters~(\ref{eq:mu_i}) and~(\ref{eq:lambda_i})
more transparent, we quantify the overall amount of bilinear $R$-parity violation
by $\xi \equiv \sqrt{\sum_i\, |\xi_i|^2}\, $ (defined such that $\sum_i |\mu_i|^2 = \mu^2 \xi^2$)
and write
\begin{equation}
  |\mu_0| = \mu \sqrt{1 - \xi^2}\, \simeq \mu\, ,  \qquad  \mu_i = c_i \mu\, \xi\, , \qquad
    \lambda_0 \simeq h_d\, \frac{\mu}{f}\, ,  \qquad  \lambda_i = d_i \frac{\mu}{f}\, \xi\, ,
\label{eq:W_parameters}
\end{equation}
where the coefficients $c_i \equiv \sum_j R^e_{ij}\, \xi_j / \xi$ and
$d_i \equiv l c_i + (l - h_d) \sum_j R^e_{ij}\, \frac{v^*_j}{v_d}\, \frac{\hat \mu_0}{\mu \xi}$
satisfy $\sum_i |c_i|^2 = 1$ and $|d_i| \sim 1$, respectively.
Eq.~(\ref{eq:W_parameters}) implies the order-of-magnitude relations $\lambda_i / \lambda_0 \sim \mu_i / \mu_0$
and, since $\xi \ll 1$ and $\mu \ll f$, the hierarchies $\mu_i \ll \mu_0$ and $\lambda_i \ll \lambda_0 \ll 1$.

Let us now determine the VEVs ($v_0$, $v_i$) and the misalignment parameters $\xi_i$. To do this,
we must minimize the full Higgs and slepton scalar potential, including the soft terms that
violate $R$-parity. The most general bilinear soft terms for the Higgs and slepton doublets
$h_u$, $h_d$ and $\tilde L_i$ are given by (prior to the filed redefinition~(\ref{eq:SU4_rotation})):
\begin{equation}
  V_{\rm soft}\, \ni\, m^2_{h_u} h^\dagger_u h_u + \hat {\tilde m}^2_{00} h^\dagger_d h_d
    + \hat {\tilde m}^2_{ij} \tilde L^\dagger_i \tilde L_j + \left( \hat {\tilde m}^2_{0i} h^\dagger_d \tilde L_i
    + \hat B_0 h_u h_d + \hat B_i h_u \tilde L_i+ \mbox{h.c.} \right) .
\label{eq:V_soft}
\end{equation}
In order to comply with the strong experimental limits on lepton flavour violating processes like
$\mu \to e \gamma$ or $\tau \to 3 \mu$, we require supersymmetry breaking to generate
close to flavour-blind slepton soft masses, i.e.
$\hat {\tilde m}^2_{ij} = {\tilde m}^2_L \delta_{ij} + \delta \hat {\tilde m}^2_{ij}$, with
$\delta \hat {\tilde m}^2_{ij} \ll {\tilde m}^2_L$.
This can be achieved e.g. by gauge-mediated supersymmetry breaking,
with the small non-universal terms $\delta \hat {\tilde m}^2_{ij}$ arising from other sources
of supersymmetry breaking  and from renormalization group running (which may split
significantly the diagonal entries of $\hat {\tilde m}^2_{ij}$).
The last three terms in Eq.~(\ref{eq:V_soft}) are not invariant under the $U(1)$ symmetry
and must arise from the decoupling of the heavy fields of mass $M$,
which in addition to the non-renormalizable superpotential operators~(\ref{eq:W_Phi})
induce terms of the form $h^\dagger_d L_i (\Phi/M)^{l-h_d}$, $h_u h_d (\Phi/M)^{h_d}$
and $h_u \tilde L_i (\Phi/M)^l$ in $V_{\rm soft}$. This results in a suppression of the
soft parameters $\hat {\tilde m}^2_{0i}$, $\hat B_0$ and $\hat B_i$
by $\epsilon^{l-h_d}$, $\epsilon^{h_d}$ and $\epsilon^l$, respectively.
We thus have
\begin{equation}
  \hat {\tilde m}^2_{ij}\, =\, {\tilde m}^2_L \delta_{ij} + \delta \hat {\tilde m}^2_{ij}\, ,
  \qquad \hat {\tilde m}^2_{0i}\, \sim\, {\tilde m}^2_L\, \epsilon^{l-h_d}\, ,
  \qquad \hat B_i\, \sim\, \hat B_0\, \epsilon^{l-h_d}\, .
\label{eq:oom_relations}
\end{equation}
These relations, together with $\hat \mu_i \sim \hat \mu_0\, \epsilon^{l-h_d}$
(see Eq.~(\ref{eq:hierarchy_initial_parameters})), are enough to ensure $v_i \ll v_0$. Minimizing
the scalar potential, one obtains, at leading order in the small parameters $\hat \mu_i$, $\hat B_i$,
$\hat {\tilde m}^2_{0i}$ and $\delta \hat {\tilde m}^2_{ij}$ ($i \neq j$):
\begin{equation}
  \frac{v_i}{v_0}\ =\ \frac{\hat {\tilde m}^2_0}{\hat {\tilde m}^2_i} \left( \frac{\hat B_i}{\hat B_0}
    - \frac{\hat {\tilde m}^2_{0i} + \hat \mu_i \hat \mu^*_0}{\hat {\tilde m}^2_0} \right)^{\! *}\, ,
\end{equation}
\begin{equation}
  \xi^2\, =\, \sum_i\, \left|\, \frac{\hat {\tilde m}^2_0}{\hat {\tilde m}^2_i} \left( \frac{\hat B_i}{\hat B_0}
    - \frac{\hat {\tilde m}^2_{0i} + \hat \mu_i \hat \mu^*_0}{\hat {\tilde m}^2_0} \right)
    - \frac{\hat \mu_i}{\hat \mu_0}\, \right|^2 ,
\end{equation}
where $\hat{\tilde m}^2_0 \equiv \hat{\tilde m}^2_{00} + |\hat \mu_0|^2 + \frac{g^2 + g^{\prime 2}}{4}\, (v^2_d - v^2_u)$
and $\hat{\tilde m}^2_i \equiv \hat{\tilde m}^2_{ii} + \frac{g^2 + g^{\prime 2}}{4}\, (v^2_d - v^2_u)$.
Using Eq.~(\ref{eq:oom_relations}) and $\hat \mu_i \sim \hat \mu_0\, \epsilon^{l-h_d}$, one finally
arrives at the order-of-magnitude estimates
\begin{equation}
  \frac{v_i}{v_0}\ \sim\ \epsilon^{l-h_d}\, ,  \qquad  \xi\ \sim\ \epsilon^{l-h_d}\, .
\label{eq:alignment}
\end{equation}
The hierarchy of vevs $v_i \ll v_0$ and the alignment condition $\xi \ll 1$ are therefore a direct
consequence of the $U(1)$ symmetry. Finally, Eqs.~(\ref{eq:W_parameters}) and~(\ref{eq:alignment}) imply
\begin{equation}
  \mu_i\, \sim\, \mu_0\, \epsilon^{l-h_d}\, ,  \qquad  \lambda_i\, \sim\, \lambda_0\, \epsilon^{l-h_d}\, ,
\end{equation}
while the upper bounds $\lambda_{ijk} \lesssim \epsilon^l$, $\lambda'_{ijk} \lesssim \epsilon^l$
can be inferred from Eqs.~(\ref{eq:initial_parameters}) and~(\ref{eq:SU4_rotation}).

\section{Constraints on the pseudo-Nambu-Goldstone boson}    %
\label{app:PNGB}                                                                          %

\renewcommand{\theequation}{B.\arabic{equation}}
\setcounter{equation}{0}  

In this appendix, we discuss the experimental constraints on the mass and couplings of the
pseudo-Nambu-Goldstone boson (PNGB) $a$ associated with the spontaneous breaking
of the global $U(1)$ symmetry.
While its supersymmetric partners, the CP-even scalar $s$ and the pseudo-Goldstone fermion $\chi$,
get their masses mainly from supersymmetry breaking, the mass of the PNGB $a$ is solely due to
the sources of explicit breaking of the global $U(1)$ symmetry, assumed to be small. Hence, $a$
is the only light scalar in the model and various constraints from particle physics experiments
and astrophysical observations apply to it.

The couplings of the PNGB to photons and to electrons are the most severely constrained.
The first one is induced by the anomaly of the global $U(1)$ symmetry and is given by\footnote{The scale
$f$ that appears in Eqs.~(\ref{eq:agg}) to~(\ref{eq:Zchichi_achichi})
differs by a factor $\sqrt{2}$ from the one introduced in Section~\ref{sec:model}. Namely, we use
in this Appendix the normalization $\langle \Phi \rangle \equiv f / \sqrt{2}$, such that
the PNGB transforms as $a \to a + \alpha f$ under the global $U(1)$ symmetry.
As a consequence, the value of $f$ corresponding to the example point with $m_\chi = 110 \GeV$
in Table~\ref{tab:example_point} is therefore $f = \sqrt{2}\, (15.1 \TeV) = 21.3 \TeV$.}
\begin{equation}
\mathcal{L}_{a \gamma \gamma} = - \frac{G_{a \gamma \gamma}}{4}\, a\, F^{\mu \nu} \tilde F_{\mu\nu}\, ,
  \qquad  G_{a \gamma \gamma} = - \frac{e^2 h_d}{8 \pi^2 f}\, ,
\label{eq:agg}
\end{equation}
where $F_{\mu \nu}$ is the electromagnetic field strength, and only the down-type higgsino contributes
to the anomaly, since the charged leptons have vector-like $U(1)$ charges.
Using $h_d = 1$ and $f = 21.3 \TeV$, we obtain $|G_{a \gamma \gamma}| = 5.45 \times 10^{-8} \GeV^{-1}$
and $\tau_{a \gamma \gamma} \equiv 1 / \Gamma (a \to \gamma \gamma) = 64 \pi / (|G_{a \gamma \gamma}|^2 m^3_a)
= 4.45 \times 10^{-8}\, \mbox{s}  \left( 1 \GeV / m_a \right)^3$.
For $m_a > \mbox{few}\, 10 \MeV$, we end up in the region of the ($m_a$, $\tau_{a \gamma \gamma}$) parameter 
space where the PNGB decays before neutrino decoupling~\cite{Cadamuro:2011fd,Millea:2015qra},
so that it does not affect cosmological observables (mainly the cosmic microwave background
and the primordial light element abundances from Big Bang nucleosynthesis).
Requiring $m_a \geq 400 \MeV$, we also evade astrophysical bounds (the most stringent one
being from Supernovae 1987A) and constraints from beam dump experiments,
as can be seen from Fig.~4 of Ref.~\cite{Bauer:2017ris} (in which
$|C_{\gamma \gamma} / \Lambda| = |G_{a \gamma \gamma}| / (16 \pi \alpha) \simeq 2.7 |G_{a \gamma \gamma}|$).
The model also predicts couplings of the form $a \gamma Z$, $a Z Z$ and $a W W$,
with coefficients of similar magnitude to $G_{a \gamma \gamma}$,
but they are much less constrained~\cite{Bauer:2017ris}.

Due to its Goldstone nature, the interactions of the PNGB $a$ with matter fields are of the form
$(\partial_{\mu}a / f) J^{\mu}$, where $J^{\mu}$ is the $U(1)$ current. Its tree-level couplings
to the fermions are thus (using 2-component spinor notations for the chiral fermions $\nu_\alpha$,
$\tilde h^0_d$, $\tilde h^-_d$, $\chi$)
\begin{equation}
  \mathcal{L}^{\rm tree}_{a ff}\, =\, \frac{\partial_\mu a}{f}
    \left\{ l \left( \bar e_\alpha \gamma^\mu e_\alpha + \bar \nu_\alpha \bar \sigma^\mu \nu_\alpha \right)
    + h_d \left( \bar{\tilde h}^0_d\,\bar \sigma^\mu \tilde h^0_d + \bar{\tilde h}^-_d\, \bar \sigma^\mu \tilde h^-_d \right)
    - \bar \chi\, \bar \sigma^\mu \chi  \right\} .
\label{eq:aff_tree}
\end{equation}
The last term in Eq.~(\ref{eq:aff_tree}) comes from the kinetic term $\int\! d^4 \theta\, \Phi^\dagger \Phi$
of the superfield $\Phi$ whose VEV breaks the global $U(1)$ symmetry~\cite{Bellazzini:2011et}
(working in the parametrization $\Phi = (f / \sqrt{2})\, e^{- \sqrt{2}\, G/f}$,
$G = (s+ia) / \sqrt{2} + \sqrt{2}\, \theta \chi + \theta^2 F$, in which $a$ has the shift symmetry $a \to a + \alpha f$).
Since the charged leptons have vector-like and generation-independent $U(1)$ charges, their couplings
to the PNGB vanish on shell (as can be shown by integrating by parts the terms
$l (\partial_\mu a / f)\, \bar e_\alpha \gamma^\mu e_\alpha$ in Eq.~(\ref{eq:aff_tree})),
thus evading the strong bounds from red giant cooling~\cite{Viaux:2013lha}.
Generation-dependent charges would induce off-diagonal couplings of the form
$g_{\alpha \beta} (a / f)\, \bar e_\alpha e_\beta$, which would mediate flavour-changing
processes such as $\mu \to e\, a$~\cite{Wilczek:1982rv}.
Finally, since the model is supersymmetric, the PNGB also couples to the charged sleptons, sneutrinos
and to the scalar components of the down-type Higgs doublet, but these couplings vanish on shell.

\begin{figure}[t]
 \centering
 \includegraphics[width=0.7\textwidth]{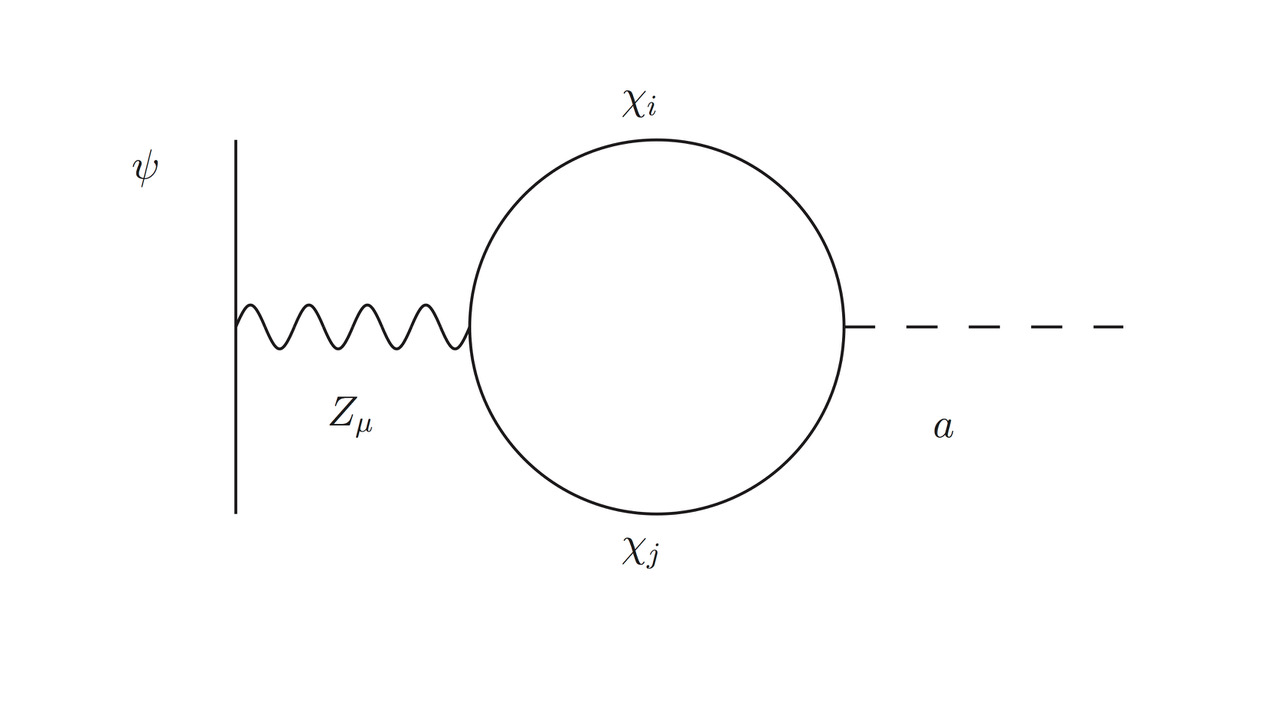}
 \caption{\small 1-loop coupling of the PNGB $a$ to a pair of fermions (Z-mediated diagram).}
 \label{fig:loopcoupl}
\end{figure}

At the one-loop level, a coupling of the PNGB to two electrons is induced by diagrams mediated
by the W and the Z bosons~\cite{Chikashige:1980ui}.
The Z-mediated diagram depicted in Fig.~\ref{fig:loopcoupl} induces the followng terms
in the Lagrangian~\cite{Feng:1997tn}
\begin{equation}
\mathcal{L}_{a\psi\psi} = - i g_{a\psi\psi}\, a\, \bar \psi \gamma^5 \psi\, ,  \qquad
  g_{a\psi\psi} = \frac{g^2}{\cos^2 \theta_W}\, \frac{I(m^2_a)}{m^2_Z - m^2_a}\, T^3_\psi\, \frac{m_\psi}{f}\, ,
\label{eq:aff_1loop}
\end{equation}
where $\psi$ is any fermion coupling to the $Z$ boson, $m_\psi$ and $T^3_\psi$ its mass and third
component of weak isospin. The loop function $I(p^2)$ is given by
\beq
  I(p^2)\, =\, \frac{1}{8 \pi^2}\ \sum_{i,j}\, \int^1_0 dx \left( \alpha_{ij} \Delta_{ij} (p^2 = 0) \ln \frac{\Delta_{ij}}{\Lambda^2}
  - \beta_{ij} M_i M_j \ln \frac{\Delta_{ij}}{\Lambda^2} \right) ,
\eeq
where the sum runs over all pairs of charginos $\{ (\tilde \chi^\pm_i, \tilde \chi^\pm_j); i, j = 1 \cdots 5 \}$
and neutralinos $\{ (\tilde \chi^0_i, \tilde \chi^0_j); i, j = 1 \cdots 8 \}$ with masses $M_i$ and $M_j$,
$\Delta_{ij} (p^2) = (1-x) M^2_i + x M^2_j - x(1-x) p^2$ and $\Lambda$ is the cutoff of the effective
field theory, which we identify with $f$. The coefficients $\alpha_{ij}$ and $\beta_{ij}$ are given by
$\alpha_{ij} = g^{L*}_{ij} q^L_{ij} + g^{R*}_{ij} q^R_{ij}$ and
$\beta_{ij} = g^{L*}_{ij} q^R_{ij} + g^{R*}_{ij} q^L_{ij}$, where $g^{L,R}_{ij}$ and $q^{L,R}_{ij}$
are the couplings of the charginos and neutralinos to the Z boson and PNGB, respectively:
\begin{equation}
  \mathcal{L}_{Z \chi_i \chi_j}\, =\, \frac{g}{\cos \theta_W}\ Z_\mu\,
    \bar \chi_i \gamma^\mu (g^L_{ij} P_L + g^R_{ij} P_R) \chi_j\, ,  \qquad
  \mathcal{L}_{a \chi_i \chi_j}\, =\, \frac{\partial_\mu a}{f}\
    \bar \chi_i \gamma^\mu (q^L_{ij} P_L + q^R_{ij} P_R) \chi_j\, .
\label{eq:Zchichi_achichi}
\end{equation}
The coefficients $g^{L,R}_{ij}$ (resp. $q^{L,R}_{ij}$) are obtained by diagonalizing the chargino
and neutralino mass matrices and writing the $Z$ boson couplings to leptons, higgsinos and
charged wino (resp. the PNGB couplings to fermions~(\ref{eq:aff_tree})) in terms of the mass
eigenstates $\chi_{i,j}$ ($= \tilde \chi^+_{i,j}$ or $\tilde \chi^0_{i,j}$).

Let us now focus on the coupling of the PNGB to electrons. Assuming $f = 21.3 \TeV$ and $m_a \lesssim 1 \GeV$,
we obtain $g_{aee} = 6.8 \times 10^{-11}$ from Eq.~(\ref{eq:aff_1loop});
the contribution of the W diagram is expected to be of the same order of magnitude.
Using the dictionary $c_{ee} / \Lambda = g_{aee} / m_e$, we can see from Fig.~4 of Ref.~\cite{Bauer:2017ris}
that an axion-like particle with $g_{aee} = {\mathcal O} (10^{-10})$ is subject to constraints from red giant cooling
and from Edelweiss.  The most stringent bound is the red giant one, $|g_{aee}| \leq 4.3 \times 10^{-13}$
($95 \%$~C.L.)~\cite{Viaux:2013lha} (a stronger prelimineray limit, $|g_{aee}| \leq 2.57 \times 10^{-13}$
($95 \%$~C.L.), has been given in Ref.~\cite{Straniero:2018fbv}).
However, these constraints do not apply if $m_a > \mbox{few}\,10 \keV$.
The Borexino limit shown on the same figure, which is valid for $m_a \leq 5 \MeV$,
assumes a specific axion model. One should consider instead the model-independent Borexino constraint
$|g_{aee} g_{3aNN}| \leq 5.5 \times 10^{-13}$ ($90 \%$~C.L.)~\cite{Bellini:2012kz}, where
$g_{3aNN} = (g_{ann} - g_{app})/2$ is the so-called isovector axion-nucleon coupling.
Since quarks are not charged under the $U(1)$ symmetry of Section~\ref{sec:model},
$g_{ann}$ and $g_{app}$ arise at the one-loop level, thus significantly weakening the constraint
on $g_{aee}$\footnote{One can estimate $g_{3aNN} \sim (m_N/m_e)\, g_{aee} \approx 2000\, g_{aee}$
(where $m_N$ is the nucleon mass), from which the Borexino limit on $|g_{aee} g_{3aNN}|$
can be converted into the approximate upper bound $|g_{aee}| \lesssim 10^{-8}$.}.
Furthermore, this bound does not apply for $m_a > 5 \MeV$.

In summary, the mass of the PNGB associated with the $U(1)$ symmetry of Section~\ref{sec:model} is constrained to be larger
than about $400 \MeV$ by cosmology, astrophysics and beam dump experiments.


\end{document}